\documentstyle[12pt,citesort]{article}

\catcode`\@=11

\newcount\hour
\newcount\minute
\newtoks\amorpm \hour=\time\divide\hour by 60\minute
=\time{\multiply\hour by 60 \global\advance\minute by-\hour}
\edef\standardtime{{\ifnum\hour<12 \global\amorpm={am}%
        \else\global\amorpm={pm}\advance\hour by-12 \fi
        \ifnum\hour=0 \hour=12 \fi
        \number\hour:\ifnum\minute<10
        0\fi\number\minute\the\amorpm}}
\edef\militarytime{\number\hour:\ifnum\minute<10
0\fi\number\minute}

\def\draftlabel#1{{\@bsphack\if@filesw {\let\thepage\relax
   \xdef\@gtempa{\write\@auxout{\string
      \newlabel{#1}{{\@currentlabel}{\thepage}}}}}\@gtempa
   \if@nobreak \ifvmode\nobreak\fi\fi\fi\@esphack}
        \gdef\@eqnlabel{#1}}
\def\@eqnlabel{}
\def\@vacuum{}
\def\marginnote#1{}
\def\draftmarginnote#1{\marginpar{\raggedright\scriptsize\tt#1}}
 \def \lc {light-cone\ }

\def\draft{
        \pagestyle{plain}
        \overfullrule=2pt
        \oddsidemargin -.5truein
        \def\@oddhead{\sl \phantom{\today\quad\militarytime} \hfil
        \smash{\Large\sl DRAFT} \hfil \today\quad\militarytime}
        \let\@evenhead\@oddhead
        \let\label=\draftlabel
        \let\marginnote=\draftmarginnote
        \def\ps@empty{\let\@mkboth\@gobbletwo
        \def\@oddfoot{\hfil \smash{\Large\sl DRAFT} \hfil}
        \let\@evenfoot\@oddhead}
        \def\@eqnnum{(\theequation)\rlap{\kern\marginparsep\tt\@eqnlabel}%
        \global\let\@eqnlabel\@vacuum}  }

\newcommand{\rf}[1]{(\ref{#1})}
\renewcommand{\theequation}{\thesection.\arabic{equation}}
\renewcommand{\thefootnote}{\fnsymbol{footnote}}

\newcommand{\newsection}{    
\setcounter{equation}{0}
\section}
\def\appendix#1{
  \addtocounter{section}{1}
  \setcounter{equation}{0}
  \renewcommand{\thesection}{\Alph{section}}
  \section*{Appendix \thesection\protect\indent \parbox[t]{11.65cm}
  {#1} }
  \addcontentsline{toc}{section}{Appendix \thesection\ \ \ #1}
  }

\def\apr{{{\rm A}^\prime}}
\def\bpr{{{\rm B}^\prime}}
\def\cpr{{{\rm C}^\prime}}

\def\sca{{\scriptscriptstyle{\cal  A}}}
\def\scb{{\scriptscriptstyle{\cal  B}}}

\def\Csp{C^\prime}

\def\vm{{\mu}}
\def\vn{{\nu}}

\def\ssmP{{\scriptscriptstyle P}}
\def\ssmK{{\scriptscriptstyle K}}
\def\ssmD{{\scriptscriptstyle D}}
\def\ssmQ{{\scriptscriptstyle Q}}
\def\ssmS{{\scriptscriptstyle S}}

\def \td {\tilde }

\def \gg {{\cal g}}

\def \foot {\footnote}
\def \bi{\bibitem}
\def \la {\label}

\def \ha {{1 \over 2}}

\def \ov {\over}

\def\nline{\,\nabla\kern -0.7em\raise0.2ex\hbox{/}\,\,}
\def\yline{\,y\kern -0.47em /}
\def\aline{\,a\kern -0.49em /}
\def\parline{\,\partial\kern -0.55em /\,\,}

\def \s{\sigma}

\def \del{\partial}

\def\det{\hbox{det}}
\def\be{\begin{equation}}
\def\ee{\end{equation}}

\def \ci {\cite}

\def \G {\Gamma}
\def \k {\kappa}

\def \L {{\cal L}}

\def\apr{{A'}}

\def \n {\nu}

\def\x'{\mathaccent 19 x}
\def\y'{\mathaccent 19 y}
\def\n'{\mathaccent 19 n}
\def\u'{\mathaccent 19 u}

\def\X'{\mathaccent 19 X}
\def\Y'{\mathaccent 19 Y}
\def\Z'{\mathaccent 19 Z}

\def\et'{\mathaccent 19 \eta}
\def\th'{\mathaccent 19 \theta}
\def\lam'{\mathaccent 19 \lambda}
\def\varet'{\mathaccent 19 \vartheta}
\def\rh'{\mathaccent 19 \rho}
\def\ph'{\mathaccent 19 \phi}
\def\xb'{\mathaccent 19 {\bar{x}}}

\begin{document}

\begin{titlepage}
\begin{flushright}
FIAN/TD/00-18
\\
OHSTPY-HEP-T-00-029
\\
hep-th/0011191\\
\end{flushright}
\vspace{.5cm}

\begin{center}
{\LARGE
Superparticle and superstring in  $AdS_3\times S^3 $ 
Ramond-Ramond background \\[.2cm]
in light-cone gauge }
 \\[.2cm]
\vspace{1.1cm}
{\large R.R. Metsaev,${}^{{\rm a,b,}}$\footnote{\
E-mail: metsaev@lpi.ru, metsaev@pacific.mps.ohio-state.edu}
and A.A. Tseytlin${}^{{\rm a,}}$\footnote{\ Also at 
Imperial College, London and   Lebedev Institute, Moscow.\
 E-mail: tseytlin.1@osu.edu} }

\vspace{18pt}
 ${}^{{\rm a\ }}${\it
 Department of Physics,
The Ohio State University  \\
Columbus, OH 43210-1106, USA\\
}

\vspace{6pt}

${}^{{\rm b\ }}${\it
Department of Theoretical Physics, P.N. Lebedev Physical
Institute,\\ Leninsky prospect 53,  Moscow 117924, Russia
}

\end{center}

\vspace{2cm}

\begin{abstract}

We discuss  superparticle and superstring
 dynamics
in $AdS_3\times S^3$ supported by R-R 3-form background
using light-cone gauge approach.
Starting with  the   superalgebra
$psu(1,1|2)\oplus\widetilde{psu}(1,1|2)$
representing the basic  symmetry of this background 
we find the light-cone superparticle Hamiltonian.
We determine the harmonic decomposition of 
light-cone superfield describing  fluctuations of type 
IIB supergravity fields expanded near 
 $AdS_3\times S^3$ background and thus the 
 corresponding Kaluza-Klein 
spectrum.  We fix the fermionic and bosonic 
light-cone
gauges in the covariant 
Green-Schwarz 
$AdS_3 \times S^3$ superstring action 
and find the corresponding 
light-cone  string Hamiltonian.
We also obtain a  realization of the generators  of 
$psu(1,1|2)\oplus \widetilde{psu}(1,1|2)$ in terms of  
the superstring  2-d fields in the light-cone gauge.
\end{abstract}

\end{titlepage}
\setcounter{page}{1}
\renewcommand{\thefootnote}{\arabic{footnote}}
\setcounter{footnote}{0}

\def \adss {$AdS_5 \times S^5$\ }
\def \N {{\cal N}}
\def \lc {light-cone\ }
\def \ta { \tau}
\def \s { \sigma }
\def  \gg  { {\rm g}}
\def \sg {\sqrt {g }}
\def \te {\theta}
\def \vp {\varphi}
\def \xp {x^+}
\def \xm {x^-}
\def \p {\phi}
\def \vt {\theta}
\def \a { \alpha}
\def \r {\rho}
\def \fourth {{1 \ov 4}}
\def \DD {{\cal D}}
\def \half {{1 \ov 2}}
\def \inv {^{-1}}
\def \D {{\cal D}}
\def \DD {{\rm D}}
\def \vr {\varrho}
\def \diag {{\rm diag}} \def \td { \tilde }
\def \tta {\td \eta}
\def \cA {{\cal A}}
\def \cB   {{\cal B}}
\def \na {\nabla}
\def \PP {{\cal P}} 

\newsection{Introduction}

Understanding how to quantize superstrings in Ramond-Ramond 
backgrounds is of topical interest, in particular, 
in connection with string theory -- gauge theory duality
\ci{Pol,mald}. 
The basic example  of  type IIB  Green-Schwarz 
string 
in  $AdS_5\times S^5$ with  R-R  5-form  background \ci{mald}
was studied, e.g., in 
\cite{MT1,KRR,Pes,KR,KT,For,DGT,pess}.
One may hope that a  progress towards understanding 
the spectrum of this theory may be achieved by using a
light-cone gauge approach recently developed in 
\ci{MT3,MTT} (for an alternative covariant approach see \ci{berk}).
To get a better understanding 
of this  light-cone approach it may be useful 
to consider first  a similar but somewhat 
  simpler string  model. 

An obvious candidate for such simpler model 
is  type IIB string in 
$AdS_3\times S^3\times T^4$  with 
 R-R  3-form  background.\foot{In what follows
we shall ignore  the trivial $T^4$ factor.}
The    $AdS_3\times S^3\times T^4$  with 
 NS-NS  3-form  background
 represents 
  the near-horizon limit 
of NS 5-brane -- fundamental string configuration 
 \ci{cvet}
and  a fundamental superstring probe in it may be  
described  by  the standard  
$SL(2,R) \times SU(2)$ WZW model in the 
NSR formulation.
 However,  the  superstring propagation 
 in  S-dual R-R  background 
 which is   the near-horizon limit
of D5--D1 system \ci{calmal} cannot be  studied 
 directly in 
the usual  NSR formalism. 
The explicit form of the covariant GS string action in
 this R-R background
was found   in \ci{ad,st,ri} by 
 applying  the  same  supercoset method 
 which was  used in the  $AdS_5\times S^5$ string 
 case in  \ci{MT1}.
An alternative ``hybrid" 
approach  to quantization of 
superstring in $AdS_3\times S^3$ R-R background
 was developed in 
\ci{vaf} (see also \ci{ber}).

In this paper we shall discuss several 
   aspects of 
superstring dynamics in the $AdS_3\times S^3$  R-R background
in
 the light-cone approach developed for  
 the  $AdS_5 \times
S^5$ case in  \ci{MT3,MTT}.
Since the simplest limiting case  of  superstring 
is superparticle,  we also consider 
in some detail the light-cone superparticle theory 
in $AdS_3\times S^3$,  following closely the treatment 
of the $AdS_5 \times S^5$ case in \cite{met3}.
First quantization of superparticle determines 
 the spectrum of fluctuations  of type IIB 
 supergravity in $AdS_3\times S^3\times T^4$ (found directly 
 in component form in \ci{deg}) 
 and thus also the ``ground state" 
 spectrum 
 of the corresponding string
  theory.
In the treatment of the superstring theory
  our starting point will be the covariant  GS 
  action (see \ci{ad,st,ri}) where we shall fix the 
  light-cone-type  fermionic ($\k$-symmetry) and 
  bosonic (2-d diffeomorphism)  gauges 
  and derive the light-cone Hamiltonian along the lines of 
  the phase space approach of \ci{MTT}.

The paper is organized as follows.

In Section 2 we    review the structure of the  
underlying symmetry  superalgebra 
of the  type IIB superstring theory in 
$AdS_3\times S^3$  R-R background  --
$psu(1,1|2)\oplus\widetilde{psu}(1,1|2)$ \ 
\ci{clt} and present its (anti)commutation relations 
in a light-cone basis.

In Section 3 we consider  
superparticle dynamics in $AdS_3\times S^3$.
We find the light-cone 
superparticle Hamiltonian and a realization 
of the generators  of 
$psu(1,1|2)\oplus \widetilde{psu}(1,1|2)$ 
on  phase space of (first-quantized)   superparticle.

In Section 4 we develop a manifestly supersymmetric 
light-cone gauge formulation 
of type IIB supergravity  on 
$AdS_3\times S^3$ background.
 The quadratic term in the action 
 for  fluctuation fields  is written  in 
  terms of a single  unconstrained 
scalar light-cone superfield, 
 allowing us to treat all
the  component fields on an  equal footing.
We also present  a  superfield 
version of 
 $S^3$ harmonic decomposition and find the corresponding K-K  
spectra  of the supergravity 
modes propagating  in $AdS_3$. 

In Section 5 we find the 
 $\k$-symmetry light-cone gauge fixed form of the superstring
 action
in $AdS_3\times S^3$. We give  the superstring Lagrangian 
both in  the ``Wess-Zumino"
 and ``Killing"  parametrizations of the  basic
  coset 
 superspace
  
 \noindent
   $[PSU(1,1|2)\times \widetilde{PSU}(1,1|2)]/[SO(2,1)\times SO(3)]$
   on which the superstring  is 
   propagating.
We also  discuss   a reformulation of
the resulting 
 superstring  action  in terms of 2-d
Dirac world-sheet fermions.

Section 6   is devoted to 
 the  light-cone phase space approach to 
 superstring theory. We  fix the analog of the GGRT 
bosonic light-cone gauge  and derive  the 
phase space analog of the  superstring Lagrangian 
of  Section 5
and the corresponding light-cone gauge  Hamiltonian.

In Section 7 we obtain a realization of the generators  of the 
symmetry 
 superalgebra 
$psu(1,1|2)\oplus \widetilde{psu}(1,1|2)$ as 
Noether charges expressed 
 in terms of  the 2-d fields which are 
the coordinates of the  $AdS_3 \times S^3$ superstring 
 in the light cone gauge.

Some  technical details are collected 
in five Appendices.
In Appendix A we summarize our  notation  and definitions
used in this paper and 
give  some  relations relevant for a  coset description of
$S^3$.
In Appendix B we describe correspondence  between
the ``covariant" and ``light-cone" 
 forms of the  
$psu(1,1|2)\oplus \widetilde{psu}(1,1|2)$ superalgebra.
In Appendix C  we  explain 
 the construction of  Poincar\'e supercharges 
in the case of  superparticle.
In Appendix D  we  give  some details of computation
 of the spectrum  of type 
IIB supergravity fluctuations 
in $AdS_3\times S^3$.
In appendix E we present the  expressions  for the   
supercoset Cartan 1-forms which are the basic 
elements in  the construction of the GS  superstring action, 
and describe our   procedure of 
  fixing  the fermionic light-cone 
gauge in the string action.

\newsection{$psu(1,1|2)\oplus\widetilde{psu}(1,1|2)$ superalgebra}

The symmetry algebra of the $AdS_3 \times S^3$ with 
 R-R 3-form 
background  may be represented as a  
 direct sum of
  two copies of $psu(1,1|2)$ superalgebra, 
 i.e. as 
$psu(1,1|2)\oplus \widetilde{psu}(1,1|2)$ superalgebra \ci{clt}.
  The even part  of
this superalgebra consists of the  bosonic subalgebras  $su(1,1)$, $su(2)$
and $\widetilde{su}(1,1)$, $\widetilde{su}(2)$ respectively. 
$su(1,1)$ and
$\widetilde{su}(1,1)$ combine into  $so(2,2)$ algebra
  while $su(2)$ and
$\widetilde{su}(2)$  form $so(4)$ algebra.  
These $so(2,2)$ and $so(4)$ algebras
are the isometry algebras of the $AdS_3$ and $S^3$ 
factors respectively. 
The 
odd part of the superalgebra consists of 16 supercharges 
which correspond to the  16
Killing spinors of $AdS_3\times S^3$ geometry.

 The
superalgebra     $psu(1,1|2)\oplus \widetilde{psu}(1,1|2)$    
 will play the central  role in our constructions. Let 
   us review its commutation relations in the two 
 forms (``covariant" and ``light-cone") 
  we are going to use. 
 In $su(1,1)\oplus su(2)$ covariant basis the   
 $psu(1,1|2)$ superalgebra  has the following generators:
$m^\alpha{}_\beta$ and $m^i{}_j$ which are generators of $su(1,1)$ 
and $su(2)$  and 8 supercharges $q^\alpha_i$,
$q^i_\alpha$  ($\alpha,\beta=1,2; \ i,j=1,2$). 
Their (anti)commutation relations have  the following  well
known form
\be
[m^\alpha{}_\beta,m^\gamma{}_\delta]
=\delta^\gamma_\beta m^\alpha{}_\delta 
-\delta^\alpha_\delta m^\gamma{}_\beta\,,
\qquad
[m^i{}_j,m^k{}_n]=\delta^k_j m^i{}_n -\delta^i_n m^k{}_j\,,
\ee
\be
[m^\alpha{}_\beta , q^k_\gamma] =-\delta_\gamma^\alpha q^k_\beta
+\frac{1}{2}\delta_\beta^\alpha q^k_\gamma\,,
\qquad
[m^i{}_j , q^k_\alpha] =\delta_j^k  q^i_\alpha
-\frac{1}{2}\delta_j^i q^k_\alpha\,,
\ee
\be
[m^i{}_j , q_k^\alpha] =-\delta_k^i q_j^\alpha
+ \frac{1}{2}\delta_j^i q_k^\alpha\,,
\qquad
[m^\alpha{}_\beta, q_k^\gamma] =\delta_\beta^\gamma q_k^\alpha
- \frac{1}{2}\delta_\beta^\alpha q_k^\gamma\,,
\ee
\be\la{dud}
\{ q^i_\alpha , q_j^\beta\} = a (\delta_j^im^\beta{}_\alpha
+\delta_\alpha^\beta m^i{}_j)\,,
\qquad\ \ 
a^2=-1\,.
\ee
We assume the following  Hermitean conjugation rules 
\be
(m^\alpha{}_\beta)^\dagger = -m^\alpha{}_\beta\ , 
\qquad
(m^i{}_j)^\dagger =m^j{}_i\ , 
\qquad
(q^\alpha_i)^\dagger = \epsilon^{\alpha\beta}q_\beta^i\ , 
\qquad
(q_\alpha^i)^\dagger = q^\beta_i\epsilon_{\beta\alpha}\ , 
\ee
where $\epsilon^{\alpha\beta}$ is the Levi-Civita tensor: 
$\epsilon_{12}=\epsilon^{12}=1$.
The $\widetilde{psu}(1,1|2)$ superalgebra 
has the same commutation relations but with the constant 
$a$ in  \rf{dud} replaced by $\td a$ ($\td a^2=-1$)
 such that its sign is opposite 
to that of  
 $a$, i.e. $a\tilde{a}=1$.
 
It will be  useful to decompose the generators 
according to their light-cone $SO(1,1)$ group 
transformation properties (we shall call this ``light-cone basis").
In the light-cone basis the generators of 
$psu(1,1|2)\oplus
 \widetilde{psu}(1,1|2)$  include 
   translations $P^\pm$, conformal boosts $K^\pm$, 
   Lorentz rotation  $J^{+-}$, dilatation
 $D$, R-symmetry  generators of $su(2)$ and $\widetilde{su}(2)$ 
$J^i{}_j$ and  $\tilde{J}^i{}_j$, Poincar\'e algebra supercharges
$Q^{\pm i}$ and conformal algebra supercharges $S^{\pm i}$.
To simplify the  notation here we use the same  type of indices for
$su(2)$ and $\widetilde{su}(2)$. 
The Hermiteant conjugation rules are
\begin{eqnarray}
\label{herrul}
&
(P^\pm)^\dagger= P^\pm\, ,
\qquad
(K^\pm)^\dagger= K^\pm\, ,
\qquad
(Q^{\pm i})^\dagger=Q^{\pm}_i\, ,
\qquad
(S^{\pm i})^\dagger=S^{\pm}_i\, ,
&
\\
&
(J^{+-})^\dagger=-J^{+-}\, ,
\quad
D^\dagger=-D\, ,
\quad
J^i{}_j^\dagger=J^j{}_i\,,
\qquad\widetilde{J}^i{}_j^\dagger=\tilde{J}^j{}_i\ . 
&
\end{eqnarray}
The anti(commutation) relations then include 
(their derivation from the above relations
is explained in Appendix B)
\be\label{lccom1}
[P^\pm, K^\mp] = D \mp J^{+-}\ , 
\ee
\be
[D,P^\pm] =-P^\pm\,,
\quad
[D,K^\pm] =K^\pm\,,
\quad
[J^{+-},P^\pm]=\pm P^\pm\,,
\quad
[J^{+-},K^\pm]=\pm K^\pm\,,
\ee
\be
[D,Q_i^\pm] =-\frac{1}{2}Q_i^\pm\,,
\quad
[D,S_i^\pm] = \frac{1}{2}S_i^\pm\,,
\quad
[J^{+-},Q_i^\pm]=\pm \frac{1}{2}Q_i^\pm\,,
\quad
[J^{+-},S_i^\pm]=\pm\frac{1}{2} S_i^\pm\,,
\ee
\be
[S^\mp_i,P^\pm]=Q^\pm_i\,,
\quad
[Q^{\mp i},K^\pm]=S^{\pm i}\,,
\quad
\{Q^{\pm i},Q^\pm_j\}= \pm P^\pm\delta_j^i\,,
\quad
\{S^{\pm i},S^\pm_j\}=\pm K^\pm\delta_j^i\,,
\ee
\be\label{lccom4}
\{Q^{+i},S^-_j\}=\frac{1}{2}(J^{+-} - D)\delta_j^i
-\tilde{J}^i{}_j\,,
\qquad
\{Q^{-i},S^+_j\}=\frac{1}{2}(J^{+-} + D)\delta_j^i
+J^i{}_j\ , 
\ee
plus   
Hermitean conjugations of the above ones. 
The remaining relations can be summarized as follows.
The supercharges $Q^-_i$, $Q^{-i}$, $S^{+i}$, $S^+_i$
transform in the (anti)fundamental 
representations of $su(2)$ -- they are rotated 
only  by $J^i{}_j$, i.e. 

\be\label{lccom5}
[J^i{}_j, Q^{- k}] 
= \delta_j^k Q^{- i}-\frac{1}{2}\delta^i_j Q^{- k}\,,
\qquad
[J^i{}_j, Q^-_k] 
= -\delta^i_k Q^-_j + \frac{1}{2}\delta^i_j Q^-_k\,,
\ee
and the same for $S^+_i$, $S^{+i}$.
The remaining supercharges 
$Q^{+i}$, $Q^+_i$, $S^{-i}$, $S^-_i$
transform in the (anti)fundamental representations 
of $\widetilde{su}(2)$ -- 
they are rotated only by $\tilde{J}^i{}_j$, i.e.  

\be\label{lccom6}
[\tilde{J}^i{}_j, Q^{+ k}] 
= \delta_j^k Q^{+ i}-\frac{1}{2}\delta^i_j Q^{+ k}\,,
\qquad
[\tilde{J}^i{}_j, Q^+_k] 
= -\delta^i_k Q^+_j + \frac{1}{2}\delta^i_j Q^+_k\,,
\ee
and the same for $S^{-i}$, $S^-_i$.
The generators  $J^i{}_j$,
$\tilde{J}^i{}_j$ satisfy the  standard relations

\be\label{lccom11}
[J^i{}_j, J^k{}_n] =\delta^k_jJ^i{}_n -\delta^i_nJ^k{}_j\,,
\qquad
[\tilde{J}^i{}_j, \tilde{J}^k{}_n] 
=\delta^k_j\tilde{J}^i{}_n -\delta^i_n\tilde{J}^k{}_j\,.
\ee

\newsection{Superparticle 
dynamics in $AdS_3\times S^3$ backrgound} 

Before discussing superstring
it is instructive   to consider  first a 
superparticle propagating in $AdS_3 \times S^3$  space.
The covariant  Brink-Schwarz $\kappa$-symmetric  action for a 
superparticle 
in $AdS_3\times S^3$ can be obtained, e.g., 
 from the superstring
action of \cite{ad,st,ri}
by taking the zero slope  limit $\alpha^\prime \rightarrow 0$.
By applying  the light-cone gauge fixing procedure 
(see  \cite{MTT} and below)
one could then  obtain the 
superparticle light-cone gauge fixed action. 
One the other hand,  there is  a method \ci{dir}
which reduces the problem of
constructing  a new  (light-cone gauge) 
dynamical system  to the problem of finding a new
solution of the commutation relations of the  defining symmetry
algebra 
(in our case  $psu(1,1|2)\oplus \widetilde{psu}(1,1|2)$).
This method of Dirac was applied to the case of superparticle 
in $AdS_5\times S^5$
in \ci{met3} (see also \cite{met4})
and here we would like to demonstrate how it  works
for the superparticle in $AdS_3\times S^3$.
Quantization of superparticle determines the 
quadratic part of the action of type IIB supergravity
expanded  near $AdS_3\times S^3$
background. 

In  the light-cone formalism the generators 
of the 
$psu(1,1|2)\oplus$ $ \widetilde{psu}(1,1|2)$ superalgebra  
can be split into the two groups:

\begin{equation}\label{kingen}
P^+, K^+,\, 
Q^{+i},\,Q^+_i,\,S^{+i},\,S^+_i,\,
D,\,J^{+-},\, J^i{}_j,\,\tilde{J}^i{}_j\,,
\end{equation}
which we  shall refer to as kinematical generators,  and
\begin{equation}\label{dyngen}
P^-,\, \,K^-\,, Q^{-i},\, Q^-_i,\,
S^{-i},\,S^-_i \ , 
\end{equation}
which we shall  refer to as dynamical generators.
The kinematical generators have positive or zero $J^{+-}$ (Lorentz) 
charges, while the dynamical generators have negative $J^{+-}$ charges.
It turns out that  in the superfield realization
 the kinematical generators taken at  $x^+=0$
are quadratic in the  physical  fields,\foot{In general, 
 they have the
structure    $ G= G_1 + x^+G_2+(x^+)^{2}G_3$ where  $G_1$  is
quadratic  but $G_2$, $G_3$ contain higher order terms  in
second-quantized fields.} while the dynamical generators receive
higher-order interaction-dependent  corrections. 
The first step
is to find  a   free (quadratic) superfield 
representation for  the generators of 
 $psu(1,1|2)\oplus\widetilde{psu}(1,1|2)$. 
 The  generators  we  obtain  below we
will be used for the   description of IIB supergravity in
$AdS_3\times S^3$ background.

 Let us explain  step by step how the method of 
\ci{dir} works in the present  case:
 First,  we introduce  a light-cone superspace on which we are going
to realize the generators of
our  superalgebra. The  superspace coordinates include the  
 position coordinates $x^\pm$, $z$ of $AdS_3$, a unit vector 
$u^M$ representing  $S^3$, and the Grassmann 
coordinates $\theta^i$, $\eta^i$. In this  parametrization the 
metric of
$AdS_3\times S^3$ is $   (M=1,2,3,4) $
\be
ds^2 = \frac{1}{z^2}(2dx^+dx^- + dz^2) + du^M du^M\,,
\qquad\ \ \ \ \ 
u^Mu^M=1\ \ 
\ee
In formulating our results we  shall  trade the  bosonic 
coordinate $x^-$ and the Grassmann  coordinates 
 $\theta^i$, $\eta^i$  
for the  bosonic momentum  $p^+$ and the 
Grassmann  momenta $\lambda_i$,
$\vartheta_i$. 

Let us start with the 
 kinematical generators and 
consider them  on the surface of the initial data $x^+=0$.
The kinematical generators which 
have positive $J^{+-}$-charge are fixed  to be 

\be\label{kingen1}
P^+=p^+\,,
\qquad
K^+ = \frac{1}{2}z^2 p^+\,,
\ee
\be
Q_i^+  = \lambda_i\,,
\quad
Q^{+i} = p^+\theta^i\, 
\qquad
S^+_i  = \frac{1}{\sqrt{2}}z\vartheta_i\ , 
\qquad
S^{+i} = \frac{1}{\sqrt{2}}zp^+\eta^i\ , 
\ee
where the  coordinates $\theta^i$, $\eta^i$ and their momenta
$\lambda_i$, $\vartheta_i$ satisfy 
 the canonical anticommutation relations

\be
\{\lambda_i, \theta^j\}=\delta_i^j\ , \ \ \ \ \ \ \ 
\qquad
\{\vartheta_i, \eta^j\}=\delta_i^j
\ee
Let us note that in 
 the language of an action based on a 
 supercoset construction the above 
parametrization of the kinematical generators corresponds
 to  special choices 
 of  (i)   coset representative and  (ii)  light-cone gauges 
 for 1-d diffeomorphism symmetry  and $\kappa$-symmetry.
 In fact, these choices may be motivated  by a simple 
 form of the resulting generators.

Once the above  generators are chosen, 
the remaining
kinematical generators which have zero $J^{+-}$-charge are fixed by
the commutation relations of the superalgebra
\begin{eqnarray}
\label{kingen6}
&&
J^{+-}
=\partial_{p^+}p^+-\frac{1}{2}\theta\lambda
-\frac{1}{2}\eta\vartheta+1\,,
\quad
D = -\partial_{p^+}p^+ +z\partial_z
+\frac{1}{2}\theta\lambda+\frac{1}{2}\eta\vartheta
-\frac{1}{2}\,,
\\
\label{kingen8}
&&
J^i{}_j
=l^i{}_j+\eta^i\vartheta_j -\frac{1}{2}\delta^i_j\eta\vartheta\,,
\qquad \ \ \  \ \widetilde{J}^i{}_j
=\tilde{l}^i{}_j+\theta^i\lambda_j 
-\frac{1}{2}\delta^i_j\theta\lambda\,\ , 
\end{eqnarray}
where $\partial_{p^+}\equiv \partial/\partial p^+$,
$\partial_z\equiv \partial/\partial z$.
The orbital parts $l^i{}_j$ and $\tilde{l}^i{}_j$ of
the  angular momenta
$J^i{}_j$ and $\tilde{J}^i{}_j$ are given by
\be\label{lqua}
l^i{}_j =\frac{1}{4}(\sigma^{MN})^i{}_j l^{MN}\,,
\qquad
\widetilde{l}^i{}_j =\frac{1}{4}(\bar{\sigma}^{MN})^i{}_j l^{MN}\,,
\ee
where the $so(4)$ orbital momentum $ l^{MN}  $ can be 
chosen as\footnote{Note that the   concrete
parametrization of the $S^3$ part is not 
very important to us as in the
case of the superparticle all the 
generators are expressed  in terms of the 
orbital part of the angular momentum.}
\be
l^{MN} = u^M \hat{\partial}^N - u^N\hat{\partial}^M\ . 
\ee
Here  $\hat{\partial}^M$ is covariant tangent derivative on $S^3$
which is by fixed by the  constraint $u^M\hat{\partial}^M=0$
and by  the  commutation relations

\be\label{vmn}
[\hat{\partial}^M,u^N]=v^{MN}\,,
\qquad
[\hat{\partial}^M,\hat{\partial}^N]=
u^M \hat{\partial}^N
-u^N \hat{\partial}^N,
\qquad
v^{MN}\equiv \delta^{MN} - u^M u^N\,.
\ee
The operator $l^i{}_j$ satisfies the following basic relation
\be
l^i{}_k l^k{}_j =\frac{1}{2}l^2\delta^i_j +l^i{}_j\,,
\ee
 where 
$l^2\equiv l^i{}_j l^j{}_i$. The same  relation is true
for $\tilde{l}^i{}_j$.
The  Hermitean conjugation rules are
\be
\lambda_i^\dagger = p^+\theta^i\,,
\quad
\!\theta^{i \dagger} = \frac{\lambda_i}{p^+}\,,
\quad
\!\vartheta_i^\dagger = p^+\eta^i\,,\ 
\quad
\!\eta^{i \dagger} = \frac{\vartheta_i}{p^+}\,,
\quad
(\partial_{p^+}p^+)^\dagger
=-\partial_{p^+}p^+ + \theta\lambda+\eta\vartheta -2.
\ee
Once the all the 
 kinematical generators  are fixed, 
  the dynamical generators are
found from the commutation relations of the 
basic superalgebra (for details see Appendix C)
\be\label{hamden}
P^-= \frac{1}{2p^+}( \partial^2_z - \frac{1}{z^2}A ) \ , 
\ee
\begin{eqnarray}
\label{q1den}
&&
Q^-_i=
\frac{1}{\sqrt{2}p^+}\Bigl(-\vartheta_i\partial_z
-\frac{1}{z}(\eta\vartheta)\vartheta_i+\frac{1}{2z}\vartheta_i
+\frac{2}{z}(\vartheta l)_i\Bigr)\ , 
\\
\label{q2den}
&&
Q^{-i}=\frac{1}{\sqrt{2}}\Bigl(\eta^i\partial_z
-\frac{1}{z}\eta^i(\eta\vartheta)+\frac{1}{2z}\eta^i
+\frac{2}{z}(l\eta)^i\Bigr)\ , 
\end{eqnarray}
\be
K^- = -\bar{S}\frac{1}{p^+}S
-\frac{1}{2p^+}(\tilde{l}^2 +2\lambda \tilde{l} \theta)\ , 
\ee
\be
S^{-i} = \theta^i S - (\tilde{l}\theta)^i\,,
\qquad
S_i^- = \lambda_i \bar{S}\frac{1}{p^+} 
-\frac{1}{p^+}(\lambda \tilde{l})_i\,,
\ee
where the operators $A$, $S$ and $\bar{S}$ are defined  by
\begin{eqnarray}
\label{adef}
&A \equiv 
X-\frac{1}{4}\,,
\qquad
X\equiv 2l^2 + 4\vartheta l\eta +(\eta\vartheta -1)^2\,,
&
\\
&
S\equiv
 -\partial_{p^+}p^+ +\frac{1}{2}z\partial_z +\theta\lambda
+\frac{1}{2}\eta\vartheta -\frac{3}{4}\ , 
\qquad
\bar{S}\equiv \partial_{p^+}p^+ - \frac{1}{2}z\partial_z 
-\frac{1}{2}\eta\vartheta +\frac{3}{4}
&
\end{eqnarray}
and we used  the notation
\be
(\vartheta l)_i \equiv \vartheta_j l^j{}_i\,,
\quad
(l \eta)^i \equiv l^i{}_j \eta^j\,,
\quad
(\lambda\tilde{l})_i \equiv \lambda_j\tilde{l}^j{}_i\ , 
\quad
(\tilde{l}\theta)^i \equiv \tilde{l}^i{}_j\eta^j\,,
\quad
(\vartheta l\eta)\equiv \vartheta_i l^i{}_j \eta^j \ , 
\ee
\be\label{defl}
l^2 \equiv l^i{}_j l^j{}_i\,,
\quad\widetilde{l}^2 \equiv \tilde{l}^i{}_j \td l^j{}_i\,,
\qquad
\eta\vartheta \equiv \eta^i\vartheta_i\,,
\qquad
\theta\lambda \equiv \theta^i\lambda_i
\ee
In the light-cone  approach the 
operator 
$P^-$ plays the role of the
 (minus) Hamiltonian of the superparticle.
The expressions for the supercharges can be rewritten as follows
\be
Q^-_i=
-\frac{1}{\sqrt{2}p^+}\Bigl(\vartheta_i\partial_z
+\frac{1}{2z}[\vartheta_i,A]\Bigr)\,,
\qquad
Q^{-i}=\frac{1}{\sqrt{2}}\Bigl(\eta^i\partial_z
+\frac{1}{2z}[\eta^i,A]\Bigr)\ . 
\ee
As in \ci{met1,met2} we shall call  $A$ in \rf{adef}
the $AdS$ mass 
operator.
This operator satisfies the following basic relation
\be
\{[\eta^i,A],[\vartheta_j,A]\}
+2[\eta^i,A]\vartheta_j+2[\vartheta_j,A]\eta^i
=-4A\delta_j^i\ , 
\ee
which is useful in  checking that  
$\{Q_i^-,Q^{-j}\}=-\delta_i^j P^-$.
Let us  note that 
 $A$ is equal to zero only for 
massless representations
which can be realized as irreducible 
representations of the conformal algebra
\cite{met5,met2}, i.e. of $so(3,2)$ 
 in  the case of $AdS_3$.\footnote{The values  of  
 this operator for various fields
are discussed in \cite{met6}.}
Below in Section 4.2  we shall demonstrate that
 $A$ is not equal to zero 
for the whole  spectrum of the $S^3 \times T^4$ 
compactification 
of type IIB supergravity to  $AdS_3$.

The  generators given above were defined
 on the initial data surface  $x^+=0$.
In general,  they have the
structure $G=G(x^+, {\cal X}(x^+))$ where 
 ${\cal X}$ stands for all of the  dynamical variables. 
Let us use the notation
\be\label{gzero}
G|_{x^+=0} \equiv  G(0,{\cal X}(x^+))\ . 
\ee
The generators $G|_{x^+=0}$ can be 
 obtained  from the above expressions by expressing the 
dynamical variables ${\cal X}$ in terms of light-cone time
variable $x^+$ using the  Hamiltonian
equations of motion which are postulated in our approach. 
The form of the 
generators for arbitrary  $x^+$, i.e. $G(x^+,{\cal X}(x^+))$,
is then found from the  conservation laws
for the charges 
\be
J^{+-} =  J^+|_{x^+=0} + x^+ P^-\,,\ \ 
\qquad
D =  D^+|_{x^+=0} + x^+ P^- \ , 
\ee
\be\label{kpxp}
K^+ = K^+|_{x^+=0} + x^+ (D|_{x^+=0} + J^{+-}|_{x^+=0})  + x^{+2}P^-\,,
\ee
\be \label{spxp}
S^+_i=S_i^+|_{x^+=0} -{\rm i}x^+ Q^-_i\,,
\qquad
S^{+i}=S^{+i}|_{x^+=0} + {\rm i}x^+ Q^{-i}\,.
\ee
The remaining generators do not have explicit dependence 
on $x^+$, i.e. they have the structure 
$G(x^+,{\cal X}(x^+)) = G(0,{\cal X}(x^+))$.

\newsection{Light-cone gauge superfield formulation of \\
type IIB supergravity on  $AdS_3\times S^3$ 
 }

In this Section we shall present
the  light-cone gauge superfield description  of type 
IIB supergravity on  $AdS_3\times S^3$ backround, 
implied by the quantization of the superparticle 
described in the previous Section.
Linearized equations of motion for fluctuations of 
supergravity fields in $AdS_3\times S^3\times K3$  background 
and the  corresponding spectrum  
 were found in  component form 
\cite{deg}. We shall use instead the light-cone superfield approach. 

 This analysis  can be viewed as 
 a step towards understanding the spectrum
 of  string theory in $AdS_3\times S^3$.
 As is well known in the case of (super)strings in flat  space,
reproducing the correct  spectrum of the 
massless modes plays an  important role in
determining  a consistent 
quantization scheme.
  The  $AdS_3\times S^3$ 
spectrum we shall find  below 
  should   be  a useful 
guiding principle  in  quantising  superstrings in
this space. In particular, 
 the operator ordering and renormalization scheme
  should be chosen  so that the ground state of the 
superstring theory in $AdS_3\times S^3$ (with R-R
 3-form background)\foot{The selection  of R-R as opposed 
 to NS-NS background is 
 pre-determined by our choice of the basic superalgebra in Section 2.}
will have the  spectrum  described below.

Finding even the quadratic part of the 
action  for fluctuations of the supergravity fields in a curved 
background
is a complicated problem.
There  are two ways of determining  spectra 
of compactifications
of the type II  supergravity. The first one  uses  oscillator
construction \cite{gunmar}. The second  one is based on the 
analysis of  equations of motion \cite{kimrom,deg}. 
In our construction of the 
 spectrum  we shall  follow the second approach. A 
new element  which  substantially 
simplifies  the  analysis
is the use of the 
 light-cone  superfield formulation.

\subsection{Quadratic light-cone superfield  action
}

 We
could in principle use the covariant superfield description of type 
IIB
supergavity \ci{HW}, starting  with linearized
expansion of superfileds,  imposing
 light-cone  gauge on fluctuations and 
 then solving the  constraints to eliminate 
non-physical degrees of freedom 
 in terms of physical ones. That would be quite tedious. 
The  light-cone
gauge method  provides a self-contained approach 
which does not rely upon existence of a 
covariant description  and which  gives a much 
shorter way to arrive to final
results. The key idea is that, as in flat space \ci{SCH},
 the  superparticle
supercharges  found  in the  previous Section provide
 realization of the 
generators of the basic  $psu(1,1|2)\oplus\widetilde{psu}(1,1|2)$
superalgebra in terms of the 
differential operators acting on the scalar 
supergravity superfield
$\Phi(x^\pm,z,u,\theta,\eta)$.
 It is convenient to Fourier transform to the 
 momentum space for
all of the coordinates except the radial $AdS_3$ coordinate 
 $z$ and  $S^3$  directions 
$u^M$.  This means 
using $p^+$, $\lambda_i$, $\vartheta_i$
instead of $x^-$, $\theta^i$, $\eta^i$
($\lambda_i$ and $\vartheta_i$ are in  the fundamental
representations of  $\widetilde{su}(2)$ and $su(2)$). 
Thus our basic superfield will be 
 $\Phi(x^+,p^+,z,u,\lambda,\vartheta)$ with the
following expansion in powers of the 
Grassmann  momenta $\lambda_i$ and
$\vartheta_i$
\begin{eqnarray}
\Phi(x^+,p^+,z,u,\lambda,\vartheta)
&=&p^+\phi
+\lambda_i\psi_1^i+\vartheta_i\psi_2^i
+(\epsilon\lambda^2)\phi_1
+\lambda_i\vartheta_j\phi_2^{ij}+(\epsilon\vartheta^2)\phi_1^*
\nonumber\\
&+&
\frac{1}{p^+}\Bigl(
(\epsilon\lambda)^i(\epsilon\vartheta^2)\psi_1^{i*}
+(\epsilon\vartheta)^i(\epsilon\lambda^2) \psi_2^{i*}\Bigr)
-\frac{1}{p^+}(\epsilon\lambda^2)(\epsilon\vartheta^2)\phi^*\,,
\end{eqnarray}
where the coefficients $\phi,\phi_1,\phi_2, \psi_1,\psi_2$
are functions of $x^+$, the momentum  $p^+$ and 
the  bosonic coordinates $z, u^M$.
We used  the notation 
\be
(\epsilon \lambda^2)
\equiv \frac{1}{2}\epsilon^{ij}
\lambda_i\lambda_j\,,
\qquad\ \ \ \ 
(\epsilon \lambda)^i
\equiv \epsilon^{ij}
\lambda_j
\ee
and the same for  $\vartheta$. 
The only constraint which this 
superfiled is to satisfy  is the reality
constraint
\be\la{coo}
\Phi(-p^+, z, u,\lambda,\vartheta)
=(p^+)^{2}\int d^2\lambda^\dagger d^2\vartheta^\dagger\ 
e^{(\lambda_i\lambda_i^\dagger+\vartheta_i\vartheta_i^\dagger)/p^+}
(\Phi(p^+,z, u,\lambda,\vartheta))^\dagger\ , 
\ee
where we  assume  the convention $(\lambda_1\lambda_2)^\dagger =
\lambda_2^\dagger \lambda_1^\dagger$.
This  reality  constraint 
implies that the component fields $\phi,\phi_n$ are related to 
 $\phi^*,\phi_n^*$ by the Hermitean conjugation rule
 for the Fourier components, i.e. 
$(\phi^*(-p^+))^*=\phi(p^+), \ (\phi_n^*(-p^+))^*=\phi_n(p^+)      $. 
Eq. \rf{coo}  leads  also to the following
selfduality condition 
\be
\phi_2^{ij}(p^+) = -\epsilon^{ik}\epsilon^{jl}\phi_2^{kl*}(-p^+)\, \ .
\ee
The light-cone action  has  the following `non-covariant form'
\be
S = \int dx^+dz dp^+ d^3 u  d^2\lambda d^2\vartheta\,
\Phi(-p^+,z,u,-\lambda,-\vartheta)\ [ p^+({\rm i}\partial_{x^+}  + P^-)]\ 
\Phi(p^+,z,u,\lambda,\vartheta)\ , 
\ee
where the Hamiltonian density ($-P^-$)
 is given by (\ref{hamden}) and $d^3 u$ stands for 
  the $S^3$ volume element,
 i.e. $ d^4 u \delta(u^M u^M-1)$.

Transforming back to the  position coordinate $x^-$
this action can be cast  into  `relativistic-invariant' form
\be\label{covact}
S = \frac{1}{2}\int d^3x d^3 u\ d^2\lambda\ d^2\vartheta\,
\Phi(x,u,-\lambda,-\vartheta)\ (\Box  - \frac{1}{z^2}A)\ 
\Phi(x,u,\lambda,\vartheta)
\ee
where $\Box$ is the flat D'Alembertian\ 
$\Box = 2\partial_{x^-}\partial_{x^+} + \partial^2_z$,
and   $d^3x\equiv dx^+dx^-dz$.

As was already mentioned above, 
the superparticle charges found  in 
Section 3 give  the representation of 
$psu(1,1|2)\oplus \widetilde{psu}(1,1|2)$
 in terms of differential operators acting on  the 
 supergravity
superfield $\Phi$. We can thus 
 write down the ``superfield-theory" (or ``second-quantized") 
 realization of 
of $psu(1,1|2)\oplus \widetilde{psu}(1,1|2)$ 
 generators
\be
\hat{G} = \int dp^+ dzd^3u\ d^2 \lambda\  d^2\vartheta\,\ 
p^+\Phi(-p^+,z,u, -\lambda,-\vartheta)\  G \ 
\Phi(p^+,z,u,\lambda,\vartheta)\,,
\ee
where $G$ indicates representation of 
$psu(1,1|2)\oplus \widetilde{psu}(1,1|2)$ superalgebra in terms of
differential operators given in previous Section.

\subsection{Harmonic decomposition of the light-cone 
superfield\\  and the spectrum }

The light-cone description given 
above provides a convenient way to analyse the 
harmonic decomposition of basic component fields
and thus the  corresponding spectra of fluctuation modes. 
A nice feature of this 
approach is that  this can be done at the level of superfields, 
i.e. in a 
manifestly supersymmetric way.
The action (\ref{covact}) gives the following
 equation of motion for the basic superfield
$\Phi$
\be
(\Box  - \frac{1}{z^2}A)\Phi=0\ .
\ee
To find the spectrum we are thus 
 to decompose  $\Phi$ into the 
eigenvectors of the $AdS$ mass operator $A$ defined in \rf{adef}.
Let us first 
 make the standard harmonic decomposition
 (we absorb the coefficients of the expansion in the `basic'
 vectors)\footnote{In this subsection
   the index $k$ is used to indicate the Kaluza-Klein modes.}
\be
\Phi = \sum_{k=0}^\infty   \Phi_k\ , 
\ee
where  $\Phi_k$ are the $so(4)$  harmonic superfields,  
satisfying, by definition, 
\be
2l^2 \Phi_k = k(k+2)\Phi_k\ . 
\ee
We can further expand  each  $\Phi_k$ in power 
series with respect to the Grassmann  momentum $\vartheta$
writing 
\be
\Phi_k = \sum_{\sigma=0}^2 \Phi_{k,\sigma}\ , \ \ \ 
\ee
where $\Phi_{k,\sigma}$ satisfies
\be
2l^2 \Phi_{k,\sigma}  =k(k+2) \Phi_{k,\sigma}\,,\ \ \ 
\qquad
\eta\vartheta \Phi_{k,\sigma} = (2-\sigma ) \Phi_{k,\sigma}\ . 
\ee
These equations tell us that the harmonic 
superfield $\Phi_{k,\sigma}$ is a polynomial
of  degree $\sigma$ in the  Grassmann 
momentum $\vartheta$. {}From the expression for the operator 
$X$ (\ref{adef}) 
it is then clear the superfields $\Phi_{k,0}$ are its 
eigenvectors 
\be
X\Phi_{k,0} = (k+1)^2\Phi_{k,0}\ . 
\ee
It is easy to demonstrate that  $\Phi_{k,2}$ are  also the 
eigenvectors of $X$ with the same eigenvalues, i.e. 
$X\Phi_{k,2} = (k+1)^2\Phi_{k,2}$.
This gives the following equations of motion
 determining the part of the 
 mass spectrum corresponding  to $\Phi_{k,0}, \ \Phi_{k,2}$
\be
\Bigl(\Box -\frac{(2k+1)(2k+3)}{4z^2}\Bigr)\Phi_{k,0}=0\,,
\qquad
\Bigl(\Box -\frac{(2k+1)(2k+3)}{4z^2}\Bigr)\Phi_{k,2}=0\,.
\ee
 It turns out that
the remaining superfields $\Phi_{k,1}$
 are  not  eigenvectors of $X$. 
They can be decomposed,  however, 
 into the  eigenvectors of this operator 
 as follows 
(for details see Appendix B)
\be
\Phi_{k,1} 
=\Phi_{k,1}^{(1)}+\Phi_{k,1}^{(2)}\ , 
\ee
where
\be
\Phi_{k,1}^{(1)} 
= (\vartheta_i -\frac{2}{k+2} (\vartheta l)_i)\Phi_{k,1}^i\,,
\qquad
k \geq 0\,;
\ee
\be
\Phi_{k,1}^{(2)} = (\vartheta_i -\frac{2}{k} (\vartheta l)_i)
\Phi_{k,1}^i\,,
\qquad
k>0\,.
\ee
Here  $\Phi_{k,1}^i$ 
does not depend on the 
Grassmann momentum $\vartheta$ but   still depend on
Grassmann momentum $\lambda$.
Then 
\be
X \Phi_{k,1}^{(1)} = k^2 \Phi_{k,1}^{(1)}\,,
\qquad
X \Phi_{k,1}^{(2)} = (k+2)^2 \Phi_{k,1}^{(2)}\,,
\ee
and this gives the following equations of motion 
\be
\Bigl(\Box -\frac{(2k-1)(2k+1)}{4z^2}\Bigr)\Phi_{k,1}^{(1)}=0\ ,
\ \ \ \ \ \     \Bigl(\Box
-\frac{(2k+3)(2k+5)}{4z^2}\Bigr)\Phi_{k,1}^{(2)}=0\ , 
\ee
determining the  spectra of these superfields.

Note that the operator $A$ is equal to zero 
(i.e. $X= { 1 \over 4}$) 
only for massless representations
which can be realized as irreducible representations of
the  conformal algebra
\cite{met2,met5} ( $so(3,2)$ in  the case of $AdS_3$). 
{}From the above spectra one can  see that the  mass terms,
 i.e. the eigenvalues of the  operator $A$, 
are never equal to zero. That means, in particular,  that 
the fluctuation  modes   for the 
 compactification of IIB supergravity 
on $S^3$ do not satisfy the conformally invariant equations of motion
in $AdS$ space.

\newsection{Light cone superstring action \\
in $AdS_3\times S^3$  R-R background }

In this Section we  shall find the form of the type IIB 
superstring action in $AdS_3\times S^3\times T^4 $ 
background with R-R 3-form flux in the light-cone gauge.
The  Green-Schwarz action for a superstring  background
was constructed in \ci{ad,st,ri} following
a  similar construction for $AdS_5
\times S^5$ case in \ci{MT1}.
Our discussion  of light-cone gauge fixing 
will also repeat  closely the same steps as in  
refs. \ci{MT3,MTT}    where the $AdS_5 \times S^5$ case was treated.

In flat space  superstring light-cone
 gauge  fixing procedure 
 in flat space  consists of the two  stages:

(I) fermionic light-cone gauge choice,
i.e. fixing the $\k$-symmetry by  $\G^+ \theta^I=0$

(II) bosonic \lc gauge choice, i.e.
using   the conformal gauge\foot{We  use Minkowski
signature 2-d world sheet  metric $g_{\vm\vn}$ with
$g\equiv - \det g_{\vm\vn}$.}
$\sqrt {g}  g^{\vm\vn} =\eta^{\vm\vn}$
and fixing  the  residual conformal diffeomorphism symmetry
by $x^+(\tau,\sigma)  =  p^+ \tau$.

Our  fermionic  $\k$-symmetry light-cone gauge  will be 
different from the naive $\G^+ \theta^I =0$ but   will be 
   related  to
it in the flat space limit. It will  reduce the 16 fermionic coordinates
$\theta^I_\a$  to 8 physical Grassmann  variables: ``linear" 
$\theta^i$  and   ``nonlinear" $\eta^i$ and their Hermitian
conjugates $\theta_i$  and $\eta_i$. As in the case of the
superparticle the 2-d fields 
 $\theta^i$, $\theta_i$ and $\eta^i$, $\eta_i$
transform  according to the fundamental representations of
$\widetilde{SU}(2)$ and $SU(2)$ respectively. The superconformal
algebra $psu(1,1|2)\oplus \widetilde{psu}(1,1|2)$   dictates that
these variables   should  be  related to  the  Poincar\'e and 
  conformal supersymmetry   in the light-cone gauge
description of the boundary theory. As in the case of
superparticle the superstring  action and symmetry generators
will have simple (quadratic) dependence on $\theta^i$, but
complicated (quartic) dependence on $\eta^i$. \footnote{Note
that it is these fermionic 
 coordinates that are most suitable for 
  light-cone gauge fixing of 
kappa symmetry in AdS space, both in the superparticle 
and superstring cases.
  These
coordinates  were  introduced in \cite{met3}  in the study of 
light-cone gauge dynamics of  superparticle in  $AdS_5\times
S^5$.  Light-cone gauge superstring action in  $AdS_5\times S^5$
written in terms of these coordinates  was found in \cite{MT3}.} The
light-cone  gauge action  can be  found  in   two related forms.
One of them corresponds to the choice of the 
 Wess-Zumino type gauge in
superspace while another is based on the  Killing gauge. These
``gauges"   or   ``parametrizations" do not
reduce the number of fermionic degrees of freedom but only
specialize a  choice  of fermionic coordinates.

\subsection{Fermionic light-cone gauge  
action\\ in WZ pa\-ra\-metri\-zation}

Let us consider  first fixing 
fermionic light-cone gauge in the 
action written  in the  WZ parametrization.
This action turns out to be 
 more convenient  for  
reformulation  of superstring action 
in terms of 
2-$d$ Dirac spinors (see next Section).
Using 
the parametrization  of the  basic supercoset
$[PSU(1,1|2)\times \widetilde{PSU}(1,1|2)]/[SO(2,1)\times SO(3)]$
described in Appendix E
and fixing a light-cone gauge 
 the $AdS_3\times S^3$ superstring
Lagrangian  can be written as
the sum of the bosonic term, term quadratic
in  fermions  and quartic
fermionic term 
\be
{\cal L}={\cal L}_{B}+ {\cal L}^{(2)}_{F} + {\cal L}^{(4)}_{F}
\label{oom}\ . 
\ee
Here
\be
{\cal L}_B = - \sqrt{g} g^{\vm\vn} \bigg[
e^{2\phi}\partial_\vm x^+ \partial_\vn x ^-
+ \frac{1}{2}\partial_\vm \phi \partial_\vn \phi
 + \frac{1}{2}e_\vm^\apr e_\vn^\apr \bigg] \,,
\label{bos}
\ee
where $e_\vm^\apr$ is the projection of the vielbein of $S^3$  which
in the special parametrization we will be  using  is given by
 
\be
 e_\vm^\apr   = -\frac{\rm i}{2}
Tr(\sigma^\apr \partial_\vm UU^{-1})
+\frac{\rm i}{2}Tr(\sigma^\apr
 \partial_\vm \tilde{U}\tilde{U}^{-1})\ , 
\ee

\be
U^i{}_j\equiv (e^y)^i{}_j\ , \ \ \widetilde{U}^i{}_j\equiv (e^{-y})^i{}_j\ , \ \
\ \
U^\dagger U=I\ , 
\quad\widetilde{U}^\dagger \tilde{U}=I\ , \label{uuu}
\ee
where the trace  is over $i,j=1,2,3$.
The matrices  $U \in SU(2)$, $\tilde{U}\in \widetilde{SU}(2)$ depends 
on 3 independent coordinates $y^\apr$

\be
y^i{}_j\equiv \frac{{\rm i}}{2}y^\apr (\sigma^\apr)^ i{}_j \ ,
\ \ \ \ \ \
(y^i{}_j)^* =-y^j{}_i\,,\  \ \ \
\
y^i{}_i=0\, ,
\label{dere}
\ee
where $\sigma^\apr$ are 3 Pauli matrices.
The quadratic part of the  fermionic    action is 

\begin{equation}\label{ff}
{\cal L}^{(2)}_{F}
=e^{2\phi}\partial_\vm x^+\Bigl[\frac{{\rm i}}{2} \sqrt{g}g^{\vm\vn}
(-\theta_i \tilde{{\cal D}}_\vn\theta^i -\eta_i {\cal D}_\vn \eta^i
+{\rm i}\eta_i e_\vn^i{}_j\eta^j)
+\epsilon^{\vm\vn}\eta^i C_{ij}^\prime\widetilde{{\cal D}}_\vn\theta^j\Bigr] +h.c.
\ . \end{equation}
The  $\epsilon^{\mu\nu}$ dependent (P-odd)  term in (\ref{ff})
came from the WZ term in the covariant GS action on the 
supercoset.
We used the following notation

\begin{equation}\label{covder}
{\cal D}\eta^i=d\eta^i-\Omega^i{}_j\eta^j\,,
\quad
{\cal D}\eta_i=d\eta_i+\eta_j\Omega^j{}_i\,,
\quad\widetilde{{\cal D}}\theta^i = d\theta^i-\tilde{\Omega}^i{}_j\theta^j\,,
\quad\widetilde{{\cal D}}\theta_i = d\theta_i+\theta_j\tilde{\Omega}^j{}_i\,,
\ee
\be
e^i{}_j \equiv (\sigma^\apr)^i{}_j e^\apr \ ,
\end{equation}
and $  {\cal D} = d \sigma^\mu{\cal D}_\mu$ , 
$e^i{}_j = d\sigma^\mu e_\mu^i{}_j $
where $\sigma^\vm=(\tau,\sigma)$ are 2-d coordinates.
${\cal D}$, $\tilde{\cal D}$ are the generalized spinor derivatives on $S^3$.
They have the structure 
${\cal D}=d+\Omega^i{}_j J^j{}_i$,
$\tilde{{\cal D}}=d+\tilde{\Omega}^i{}_j \tilde{J}^j{}_i$
and satisfy  the relation
${\cal D}^2=0$, $\tilde{{\cal D}}^2=0$. The connections 
$\Omega^i{}_j$, $\tilde{\Omega}^i{}_j$ are given by

\begin{equation}\label{gencon}
\Omega= dUU^{-1}\,,
\quad\widetilde{\Omega}= d\tilde{U}\tilde{U}^{-1}\,,
\qquad
d\Omega-\Omega\wedge \Omega=0   \ ,
\quad
d\tilde{\Omega}-\tilde{\Omega}\wedge \tilde{\Omega}=0   \ ,
\ee
and can be written  in terms of the $S^3$  spin  connection
$\omega^{\apr\bpr}$ and the  3-bein $e^\apr$ as follows

 \begin{equation}\label{gencon1}
\Omega^i{}_j
=-\frac{1}{4}(\sigma^{\apr\bpr})^i{}_j\omega^{\apr\bpr}
+\frac{\rm i}{2}(\sigma^\apr)^i{}_je^\apr   \,, 
\qquad\widetilde{\Omega}^i{}_j
=-\frac{1}{4}(\sigma^{\apr\bpr})^i{}_j\omega^{\apr\bpr}
-\frac{\rm i}{2}(\sigma^A)^i{}_je^\apr   \ .
 \end{equation}
$\Csp_{ij}$ is the constant charge conjugation matrix
of the $SO(3)$ Dirac matrix algebra (see Appendix A).
The Hermitean conjugation rules are:
$\theta_i^\dagger=\theta^i$, $\eta_i^\dagger=\eta^i$.

The quartic fermionic term in (\ref{oom})
depends  only on half of the   Grassmann variables --
 on  $\eta$  but not on $\theta$ 

\be
{\cal L}^{(4)}_{F} = 2 \sqrt{g} g^{\vm\vn}
 e^{4\phi}\partial_\vm x^+ \partial_\vn x^+  (\eta^i\eta_i)^2  
\label{qua}
\ee

 \subsection{``2-d  spinor"  form  of the  action  }

Like in  the flat space case \ci{GS}
and in the ``long string" cases  in $AdS_5\times S^5$ 
  \ci{DGT}  the resulting   action   can  then  be  put
into  the ``2-d spinor" form, where 
the   4+4  fermionic degrees of freedom  are 
organized into 2  Dirac 2-d spinors,
defined  in  {\it curved} 2-d geometry
(we shall  follow similar discussion 
in $AdS_5 \times S^5$ case  in \ci{MT3}).
Such action may be useful to establishing a 
relation to NSR formulation.

  In order to do that one needs to impose, 
addition to fermionic light-cone gauge, the  bosonic
light-cone gauge. 
Using  the following light-cone gauge \ci{POL}
\be
\la{choi}
x^+ = \tau\ , \ \ \ \ \ \
 \
\sg g^{\vm\vn} = \diag(- e^{-2\p}, e^{2\p})\ .
\ee
we can  write the kinetic term \rf{ff} as

\begin{equation}\label{ff2}
{\cal L}^{(2)}_{F}
=\frac{{\rm i}}{2}(\theta_i \tilde{{\cal D}}_0\theta^i 
+ \eta_i \tilde{{\cal D}}_0 \eta^i
-2{\rm i}\eta_i e_0^i{}_j\eta^j)
+ e^{2\phi}\eta^i C_{ij}^\prime \tilde{{\cal D}}_1\theta^j 
+h.c.\ , 
\end{equation}
where we used the relation
$
{\cal D}\eta^i =\tilde{\cal D}\eta^i - {\rm i}e^i{}_j \eta^j
$ (see \rf{gencon1}). 
Introducing a  2-d zweibein  corresponding to the metric in \rf{choi}
\be
e_\vm^m = \diag(e^{2\phi}, 1)\,,
\qquad
g_{\vm\vn} = - e^0_\vm e^0_\vn + e^1_\vm e^1_\vn \ ,
\la{gtg}
\ee
we can put  \rf{ff2}  in the 2-d form as follows

\begin{equation}\label{ff3}
e^{-1}{\cal L}^{(2)}_{F}
=-\frac{\rm i}{2}\bar{\psi}\varrho^m e_m^\vm \tilde{{\cal D}}_\mu\psi
+\frac{\rm i}{2}\bar{\psi}\psi\partial_1\phi
-\sqrt{2}\bar{\psi}_ie_0^i{}_j\varrho^-\psi^j
+h.c. \ , 
\end{equation}
where  $\varrho^m$  are 2-d Dirac matrices,

\be\varrho^0= {\rm i} \s_2,
\quad
\varrho^1 = \s_1,
\quad
\varrho^3 =\varrho^0 \varrho^1= \s_3\,,
\quad
\varrho^\pm\equiv \frac{1}{\sqrt{2}}(\varrho^3\pm \varrho^0)\,,
\la{dirr}\ee
$\bar \psi_i= (\psi^i)^\dagger \varrho^0$,\
$\bar{\psi}\psi$ stands for $\bar{\psi}_i\psi^i$,
$\psi^T$ denotes the transposition of 2-d spinor  and
$\psi$'s are related to the  original (2-d scalar) fermionic  variables
$\theta$'s and $\eta$'s   by\foot{In our notation  \ \
${\rm i} \bar \psi_i \varrho^m  \na_m\psi^i
= - {\rm i} \psi_1 ^\dagger (\na_0 - \na_1) \psi_1 -
{\rm i} \psi_2 ^\dagger (\na_0 + \na_1) \psi_2$, \ \
$\na_m= e_m^\vm \del_\vm$.}
\begin{equation}\label{combin}
\psi^i= \left(
\begin{array}{c}
\psi^i_1\\
\psi^i_2
\end{array}\right)\,,
\qquad
\psi_1^i=\frac{1}{\sqrt{2}}[\theta^i-{\rm i}({\Csp}^{-1})^{ij}\eta_j]\,,
\qquad
\psi_2^i=\frac{1}{\sqrt{2}}[\theta^i+{\rm i}({\Csp}^{-1})^{ij}\eta_j]\,.
\end{equation}
The quartic interaction term \rf{qua}
then  takes the  following form

\begin{equation}\label{qua2}
e^{-1}\L^{(4)}_{F} =
-(\bar{\psi}_i\varrho^-\psi^i)^2\ .
\end{equation}
 The total action is thus a  kind of
  $G/H$ bosonic sigma model coupled to
  a  Thirring-type  2-d
fermionic model  in  curved 2-d geometry  \rf{gtg}
(determined by the profile of the
radial function of the $AdS$ space), and
coupled to  some 2-d vector fields.
The interactions  are such that they  ensure the quantum
2-d conformal invariance
of the total model \ci{MT1}.


The mass term $ \bar{\psi}\psi\partial_1\phi$
in (\ref{ff3}) is similar to the one in \cite{DGT}
(where the  background string configuration was
non-constant only in the radial $\phi$ direction)
and  has its origin  in the
 $\epsilon^{\mu\nu }  e^{2\p} \del_\mu x^+ \del_\nu \phi
\eta^i\Csp_{ij} \theta^j$ term
appearing after $\eta \leftrightarrow \theta$ symmetrization of  the
$\epsilon^{\vm\vn }$ term in \rf{ff}
 (its `non-covariance' is thus a consequence of the choice
$x^+=\tau$).
The action is symmetric under shifting \
$
\psi^i \rightarrow \psi^i + \varrho^- \epsilon^i     , $
where $\epsilon^i$ is the 2-d  Killing  spinor. This
symmetry reflects the fact that our original action is symmetric under
shifting  $\theta^i$ by a Killing spinor on $S^3$.

Note also that the 2-d Lorentz invariance
 is  preserved  by the fermionic \lc
gauge (original GS fermions   $\vt$  are 2-d scalars)
 but is broken by our special choice
of the  bosonic gauge \rf{choi}.
The special form  of $g_{\vm\vn}$  in \rf{choi}
implies  ``non-covariant"  dependence on $\p$
in the  bosonic part of the action:
the action
\rf{bos} for  the field $\phi$
and  the  3-sphere    coordinates  $y^\apr$
has the form

\be
\L_B =  \ha   e^{-2\p} 
\dot{\phi}^2   - \ha e^{2\p} \ph'^2
+  \ha G_{\sca\scb}( e^{-2\p} \dot{y}^\sca  \dot{y}^\scb
-  e^{2\p}   \y'^\sca  \y'^\scb )  \, ,
\la{bose}
\ee
where $ G_{\sca\scb}$ is the metric of 5-sphere
($y^\sca$ 
are  coordinates of $S^3$). 
A consequence of the unusual $g_{\mu\nu}$
 gauge choice in \rf{choi} 
  compared to the standard  conformal gauge
is   that  now 
the $S^3$ part of the
action is no longer decoupled from the radial $AdS_3$
direction $\phi$.

\subsection{Fermionic light-cone 
gauge   action\\ in Killing pa\-ra\-metri\-zation}

Now let us consider the   string 
 action in the 
Killing parametrization. The action is  again 
formulated in
terms of 6 bosonic coordinates $(x^\pm,\phi,y^\sca)$ (${\cal A}$
label 3 independent coordinates of $S^3$)
in terms of which   the  metric of
$AdS_3\times S^3$ is 

\be
ds^2 =2 e^{2\phi}dx^+ dx^- +d\phi^2+ G_{\sca\scb}(y) 
dy^\sca dy^\scb \ , 
\la{ded}
\ee
   and 8 fermionic
coordinates $(\theta^i,\theta_i)$, $(\eta^i,\eta_i)$ 
  in the  fundamental representations of
$\widetilde{SU}(2)$ and $SU(2)$ respectively. In contrast to
the  WZ
parametrization,  the fermions in the  Killing parametrization
transform in the  linear representations of $SU(2)$ and
$\widetilde{SU}(2)$, and thus 
the covariant derivatives in WZ  case (\ref{covder})
here will become ordinary
derivatives.
The Lagrangian  is  given by the sum of the
``kinetic" and ``Wess-Zumino" terms  (see Appendices  A and B 
for notation) 

$$
{\cal L}= {\cal L}_{kin}+{\cal L}_{WZ}\ , 
$$
$$
{\cal L}_{kin}
=
-\sqrt{g}g^{\vm\vn}\Bigl[
e^{2\phi}\partial_\vm x^+ \partial_\vn x ^-
+\frac{1}{2}\partial_\vm \phi\partial_\vn \phi
+\frac{1}{2}G_{\sca\scb}(y) D_\vm y^\sca D_\vn y^\scb\Bigr]
$$
\begin{equation}
- \ \frac{{\rm i}}{2} \sqrt{g}g^{\vm\vn}
e^{2\phi}\partial_\vm x^+
\Bigl[\theta^i\partial_\vn \theta_i
+\theta_i\partial_\vn \theta^i
+\eta^i\partial_\vn \eta_i
+\eta_i\partial_\vn \eta^i 
+{\rm i}  e^{2\phi}\partial_\vn x^+(\eta^2)^2\Bigr]\ ,
\label{actkin3}
\end{equation}

\begin{equation}\label{actwz3}
{\cal L}_{WZ}
=\epsilon^{\vm\vn}
e^{2\phi}\partial_\vm x^+ \eta^i C_{ij}^U\partial_\vn\theta^j+h.c. \ , 
\end{equation}
where 
\begin{equation}\label{dyan}
D_\mu  y^\sca = \partial_\mu  y^\sca - 2{\rm i}\eta_i
(V^\sca)^i{}_j\eta^j e^{2\phi} \partial_\mu  x^+\,,
\qquad
C_{ij}^U \equiv U^k{}_i C_{kl}^\prime\tilde{U}^l{}_j\ . 
\ee
Here  $G_{\sca\scb}$ 
and  $(V^\sca)^i{}_j$ are the metric tensor and the 
Killing vectors of $S^3$ respectively (see Appendix A).
This  form of the superstring action  (which we shall call
``intermediate") is most convenient for  deriving  other 
 forms  which
differ in  the way one chooses the bosonic  coordinates that
parametrize  $AdS_3\times S^3$. 
For example,  a  useful form of the action is
found by introducing  a   unit 4-vector  $u^M$  defined

\be\label{uapr}
u^\apr = n^\apr \sin|y|\,,
\qquad
u^4 = - \cos|y|\ ,  
\ee
in terms of which the  $AdS_3\times S^3$ metric is 

\be
ds^2 =e^{2\phi}dx^adx^a +d\phi^2+ du^M du^M\ , 
\ \ \  \ \ \ \ \ \   u^M u^M=1 \
.
\ee
Then the string  action takes  the form

$$
{\cal L}_{kin}
=
-\sqrt{g}g^{\vm\vn}\Bigl[
e^{2\phi}\partial_\vm x^+ \partial_\vn x ^-
+\frac{1}{2}\partial_\vm \phi\partial_\vn \phi
+\frac{1}{2}D_\vm u^M D_\vn u^M \Bigr]
$$
\begin{equation}
-\frac{{\rm i}}{2} \sqrt{g}g^{\vm\vn}
e^{2\phi}\partial_\vm x^+
\Bigl[\theta^i\partial_\vn \theta_i
+\theta_i\partial_\vn \theta^i
+\eta^i\partial_\vn \eta_i
+\eta_i\partial_\vn \eta^i 
+{\rm i}  e^{2\phi}\partial_\vn x^+(\eta^2)^2\Bigr]\ ,
\label{actkin6}
\end{equation}

\begin{equation}\label{actwz6}
{\cal L}_{WZ}
=\epsilon^{\vm\vn}
e^{2\phi}\partial_\vm x^+ (\eta^i y_{ij}\partial_\vn\theta^j
+\eta_i y^{ij}\partial_\vn\theta_j)\ , 
\end{equation}
where we used the relation 
$C_{ij}^U=-{\rm i}C_{ik}^\prime u^k{}_j$, 
made the rescalings 
$\eta^i \rightarrow {\rm i}\eta^i$,
$\eta_i \rightarrow -{\rm i}\eta_i$
and introduced  the following notation
\be
u^i{}_j \equiv (\sigma^M)^i{}_j u^M\,,
\qquad
y_{ij} \equiv C_{ik}^\prime u^k{}_j\,,
\qquad
y^{ij} \equiv - u^i{}_k(C^{\prime -1})^{kj}\ , 
\ee
\be\label{rm}
D_\vm u^M = \partial_\vm u^M 
-2{\rm i}\eta_i (R^M)^i{}_j\eta^j e^{2\phi}\partial_\vm x^+ \ , \ \ \ \  \
(R^M)^i{}_j 
= - \frac{1}{2} (\sigma^{MN})^i{}_j u^N  \ . 
\ee
with  $\sigma^{MN}$  defined in (\ref{rrr}).
Note that $u^i{}_j$, $y_{ij}$, $y^{ij}$ transform in
the  fundamental
representation of $SU(2)$ with respect to the index $i$ and
in fundamental representation of $\widetilde{SU}(2)$
 with respect to the 
index $j$. They satisfy 
\be
y_{ij}^* = -y^{ij}\,,
\quad
u^i{}_j^* = \bar{u}^j{}_i\,,
\quad
\bar{u}^i{}_j \equiv (\bar{\sigma}^M)^i{}_j u^M\,,
\quad
\bar{u}^i{}_k(C^{\prime -1})^{kj}
=u^j{}_k(C^{\prime -1})^{ki}\,.
\ee
The  parametrization  based on 
 $u^M$  is the most convenient one 
for the  discussion
of  superparticle in $AdS_3\times S^3$
and  of harmonic  decomposition of  the 
light-cone superfield of type IIB supergravity into the  
Kaluza-Klein modes (see Sections 3,4). 
We shall use 
this parametrization in
the study of the light-cone 
superstring Hamiltonian  in Section 6.

The  superstring Lagrangian (\ref{actkin3}),(\ref{actwz3}) 
taken in any of its   forms  mentioned above
can be represented   in the following  way
\be\la{lagdec}
{\cal L}= {\cal L}_1+{\cal L}_2 \    ,   
\ee
\be
{\cal L}_1\label{l1}
= -h^{\vm\vn}\partial_\vm x^+\partial_\vn  x^-
+\partial_\vm x^+ A^\vm 
+\frac{1}{2}h^{\vm\vn}\partial_\vm x^+\partial_\vn  x^+ B
-\frac{1}{2}h^{\vm\vn} g_{\sca\scb}D_\vm y^\sca D_\vn y^\scb \ , 
\ee
\be\label{l3}
{\cal L}_2 = -\frac{1}{2}h^{\vm\vn}e^{-2\phi}
\partial_\vm \phi \partial_\vn \phi + T\   , 
\ee
where 
\be 
g_{\sca\scb}\equiv  e^{-2\phi}G_{\sca\scb} \ ,  \ \ \ \ \ \ \ \ \ 
D_\vm  y^\sca \equiv \partial_\vm y^{\sca} + F^\sca \partial_\vm x^+ \  ,  
\ee
and  $h^{\vm\vn}$ is defined by 
\be \la{hhh}
h^{\vm\vn}\equiv \sqrt{g}g^{\vm\vn}e^{2\phi}\,,
\qquad\ \ \ 
h^{00}h^{11}-(h^{01})^2=-e^{4\phi} \ . 
\ee
The decomposition (\ref{lagdec}) is made so that  
the functions $A^\vm$, $B$, $F^\sca$ depend  
 (i) on  the anticommuting  coordinates and their derivatives 
 with respect to
 both    $\tau$ and $\sigma$, and  
(ii) on  the  bosonic coordinates 
and their derivatives with respect to the world sheet spatial 
coordinate $\sigma$ only. 
The reason for  this   decomposition is that below 
 we shall  use  the 
 phase space description    with respect to the 
bosonic coordinates only,  i.e. we  shall not make the Legendre transformation
with respect to the fermionic coordinates.

In  the case of  the ``intermediate"  form  of the action 
(\ref{actkin3}),(\ref{actwz3}) these  functions are 
\begin{equation}\label{funA}
A^\vm
=-\frac{{\rm i}}{2}h^{\vm\vn}(\theta^i\partial_\vn \theta_i
+\eta^i\partial_\vn \eta_i)
+\epsilon^{\vm 1}e^{2\phi}\eta^i C_{ij}^U\th'^j+h.c. \ , 
\end{equation}
\be\label{funB}
B= e^{2\phi}(\eta^2)^2,
\qquad
F^\sca 
= -2{\rm i}e^{2\phi}\eta_i (V^\sca)^i{}_j\eta^j,
\qquad
T=-e^{2\phi}\x'^+ \eta^i C_{ij}^U\dot{\theta}^j +h.c.\,.
\ee

\newsection{Light cone  Hamiltonian approach \\ to 
 superstring in $AdS_3\times S^3$}
Our next task is to fix the bosonic part of the light-cone gauge.
We   shall use  the   generalization of the  
 phase space  GGRT  approach \cite{ggrt} to a curved
 AdS-type space described in \ci{MTT}, 
 fixing the
 diffeomorphisms in $AdS_3 \times S^3$ cases   by
 the {\it same}  gauge condition 
 as in flat space.
 Most of the discussion below will follow closely Ref. \ci{MTT}.

\subsection{Phase space Lagrangian}

Computing  the canonical momenta for the bosonic coordinates 
\be
\PP_a 
= \frac{\partial {\cal L}}{\partial \dot{x}^a}\,,
\qquad
\Pi
= \frac{\partial {\cal L}}{\partial \dot{\phi}}\,,
\qquad
\PP_\sca 
= \frac{\partial {\cal L}}{\partial \dot{y}^\sca}\,,
\ee
we get  from \rf{lagdec} 
\begin{eqnarray}
\label{can1}
\Pi 
&=& -h^{00}e^{-2\phi}\dot{\phi}^+  - h^{01}e^{-2\phi}\ph'^+\ , 
\\
\PP^+ 
&=& -h^{00}\dot{x}^+ -h^{01}\x'^+\ , 
\\
\PP^\sca 
&=&
-h^{00}\dot{y}^\sca
-h^{01}\y'^\sca +F^\sca \PP^+\ , 
\\
\label{can6}
\PP^- 
&=&
-h^{00}\dot{x}^- - h^{01}\x'^- +A^0 - B\PP^+ +\PP_\sca F^\sca \ . 
\end{eqnarray}
where $\PP^\pm \equiv \PP_\mp$, $\PP^\sca\equiv g^{\sca\scb}
\PP_\scb$. 
By applying the same  procedure as in the bosonic case 
we find then  the following 
phase space Lagrangian ${\cal L}= {\cal L}_1 + {\cal L}_2$
(see  \ci{MTT}) 
\begin{eqnarray}
{\cal L}_1
&=&
 \PP^+\dot{x}^- + \PP^-\dot{x}^+ + \PP_\sca \dot{y}^\sca
+\frac{1}{2h^{00}}\Bigl[2\PP^+\PP^- + 2e^{4\phi}\x'^+\x'^-
\nonumber
\\
&+&
g^{\sca\scb}\PP_\sca \PP_\scb
+e^{4\phi}g_{\sca\scb}D_1y^\sca D_1 y^\scb
+(\PP^{+2}-e^{4\phi}\x'^{+2})B - 2F^\sca \PP_\sca \PP^+\Bigr] 
\nonumber\\
\label{ll1}
&+&
\frac{h^{01}}{h^{00}}(\PP^+\x'^- + \PP^-\x'^+ +  \PP_\sca \y'^\sca)
-\frac{1}{h^{00}}(\PP^+ +h^{01}\x'^+)A^0
+\x'^+ A^1 \ , 
\\
\label{ll3}
{\cal L}_2
&= &
\Pi\dot{\phi} +\frac{1}{2h^{00}}e^{2\phi}(\Pi^2
+\ph'^2)
+\frac{h^{01}}{h^{00}}\Pi\ph'+ T \ . 
\end{eqnarray}
Next, we  impose the light-cone gauge
\be\label{lcg}
x^+ = \tau \,,\ \ \
\qquad
\PP^+=p^+ \ . 
\ee
Using  these  gauge conditions  in the 
action and integrating over $\PP^-$
we get the expression for  $h^{00}$
\be\label{h00}
h^{00} = -p^+ \  . 
\ee
Inserting this into   (\ref{ll1}), (\ref{ll3}) 
we get the following general  form of the 
phase space light-cone Lagrangian\foot{Note that 
the function
$T$ in (\ref{funB}) is equal to zero 
in the light-cone gauge \rf{lcg}.}
\begin{eqnarray}
{\cal L}_1
&=& 
\PP_\sca \dot{y}^\sca
-\frac{1}{2p^+}\Bigl(g^{\sca\scb}\PP_\sca \PP_\scb
+e^{4\phi}g_{\sca\scb}\y'^\sca \y'^\scb
+p^{+2}B - 2p^+ F^\sca \PP_\sca\Bigr)
\nonumber
\\
\label{genph1}
&-&
\frac{h^{01}}{p^+}(p^+\x'^-  +  \PP_\sca \y'^\sca)
+A^0 \ , 
\\
\label{genph3}
{\cal L}_2
&= &\Pi\dot{\phi} -\frac{1}{2p^+}e^{2\phi}(\Pi^2
+\ph'^2)
-\frac{h^{01}}{p^+}\Pi\ph' \ . 
\end{eqnarray}
 This  general form of the phase space  Lagrangian can be
specialized to  different choices of bosonic
coordinates by  using  the corresponding 
 functions $A^0$, $B$,
and $F^\sca$.
  For the ``intermediate" 
 case \rf{actkin3},\rf{actwz3}  these functions are given by
(\ref{funA}),(\ref{funB}) so that we get 
\begin{eqnarray}
{\cal L} = {\cal L}_1 + {\cal L}_2 
&= &
\Pi\dot{\phi} 
+ \PP_\sca \dot{y}^\sca
+\frac{{\rm i}}{2}p^+(\theta^i \dot{\theta}_i
+\eta^i\dot{\eta}_i+\theta_i \dot{\theta}^i+\eta_i\dot{\eta}^i)
\nonumber\\
&-&
\frac{e^{2\phi}}{2p^+}\Bigl[\Pi^2+\ph'^2
+2l^2 + G_{\sca\scb}\y'^\sca \y'^\scb
+p^{+2}(\eta^2)^2 + 4p^+\eta_i l^i{}_j \eta^j \Bigr]
\nonumber\\
&+&
e^{2\phi}(\eta^i C_{ij}^U\th'^j + \eta_i C^{ij}_U\th'_j)
\nonumber\\
&-&
\frac{h^{01}}{p^+}\Bigl[p^+\x'^-   + \Pi\ph'
+ \PP_\sca \y'^\sca
+\frac{{\rm i}}{2}p^+(\theta^i\th'_i+\eta^i\et'_i
+\theta_i\th'^i+\eta_i\et'^i)\Bigr] \   .
\end{eqnarray}
Here $C_U^{ij}= -(C_{ij}^U)^* $,  and  we  used the  
relation
\be\label{lcla0}
G^{\sca\scb}\PP_\sca \PP_\scb = 2l^2 \ ,\ \ \ 
\qquad
l^i{}_j \equiv {\rm i}(V^\sca)^i{}_j \PP_\sca\,,
\qquad
l^2\equiv l^i{}_jl^j{}_i\,. 
\ee
By applying a  coordinate transformation 
one gets the phase space Lagrangian 
corresponding to the case 
\rf{actkin6},\rf{actwz6} in which 
the $S^3$ part is parametrized  by the unit 4-vector 
$u^M$ 
\begin{eqnarray}\label{lag6}
{\cal L} 
&= &
\Pi\dot{\phi} 
+ \PP_M \dot{u}^M
+\frac{{\rm i}}{2}p^+(\theta^i \dot{\theta}_i
+\eta^i\dot{\eta}_i+\theta_i
\dot{\theta}^i+\eta_i\dot{\eta}^i)
\nonumber\\
&-&
\frac{e^{2\phi}}{2p^+}\Bigl[\Pi^2+\ph'^2
+2l^2 + \u'^M\u'^M
+p^{+2}(\eta^2)^2 + 4p^+\eta_i l^i{}_j \eta^j\Bigr]
\nonumber\\
&+&
e^{2\phi}(\eta^i y_{ij}\th'^j+\eta_i y^{ij}\th'_j )
\nonumber\\
&-&
\frac{h^{01}}{p^+}\Bigl[p^+\x'^-   + \Pi\ph'
+ \PP_M \u'^M
+\frac{{\rm i}}{2}p^+(\theta^i\th'_i+\eta^i\et'_i
+\theta_i\th'^i+\eta_i\et'^i)\Bigr] \  , 
\end{eqnarray}
where $\PP_M$ is the canonical momentum for $u^M$ and $l^i{}_j$  in
(\ref{lcla0}) has the following explicit form 
\be \label{lcla}
l^i{}_j = { {\rm i}  \over 2} (\sigma^{MN})^i{}_j u^M \PP^N \ .
\ee
Here and below $l^i{}_j$ is for the  classical orbital
momentum
(note that going to the superparticle limit, 
after the 
quantization we get $\PP^M=-{\rm i}\hat{\partial}^M$ 
and then the classical
orbital 
momentum $l^i{}_j$   (\ref{lcla}) becomes the quantum 
momentum $l^i{}_j$ 
in (\ref{lqua})). 
Taking into account the constraint  $u^M \PP^M =0$  
(see  (\ref{seccon})) we get 
\be
l^i{}_kl^k{}_j=\frac{1}{4}\PP^M\PP^M\delta^i{}_j\,, 
\ \ \ \ \ 
l^2 =\frac{1}{2}\PP^M\PP^M\ .
\ee
The above Lagrangian leads to  the 
following (minus) Hamiltonian 
\be\label{ham}
P^- =\int_0^1 d \sigma\   \PP^- \ , 
\ee
where the Hamiltonian density $-\PP^-$ is  given by 
\be\label{strham}
\PP^-
= 
-\frac{e^{2\phi}}{2p^+}\Bigl[\Pi^2+\ph'^2
+2l^2 + \u'^M \u'^M
+p^{+2}(\eta^2)^2 + 4p^+\eta_i l^i{}_j \eta^j\Bigr]
+e^{2\phi}(\eta^i y_{ij}\th'^j +\eta_i y^{ij}\th'_j)\, .
\ee
It should be supplemented by the  constraint
\be 
p^+\x'^- + \Pi\ph' + \PP_M \u'^M 
+\frac{{\rm i}}{2}p^+(\theta^i\th'_i+\eta^i\et'_i 
+\theta_i\th'^i+\eta_i\et'^i) =0\,.
\ee 
As usual,  this constraint  allows one  to express
 the non-zero modes of the 
bosonic coordinate $x^-$ in terms of the transverse physical  ones.

It is easy to see that in the particle theory limit the 
 superstring Hamiltonian  (\ref{strham})
reduces to the superparticle one  
which was found in
section 3 by applying  the direct method of constructing 
relativistic dynamics
\ci{dir} based on the symmetry algebra.
Indeed,   the  (quantum, operator-ordered)
  superparticle light-cone
Hamiltonian in (\ref{hamden}),\rf{adef}  can
be rewritten as follows 
\be\label{parham}
\PP^-
= -\frac{1}{2p^+}\Bigl[e^{\phi}\Pi e^\phi \Pi +e^{2\phi} 
(2l^2 +(p^+\eta^2-1)^2 + 4p^+\eta_i l^i{}_j \eta^j)\Bigr]\,.
\ee
 The  string
Hamiltonian
(\ref{strham}) reduces to (\ref{parham})  modulo terms 
``quantum" terms proportional to
$\eta^2$ and a constant (in string Hamiltonian we  ignore operator
ordering).
The  derivation of the light-cone string action 
from the covariant one given above 
thus provides, in the particle limit, 
also a self-contained Lagrangian 
derivation of the light-cone gauge superparticle 
Hamiltonian  (\ref{hamden}) (obtained indirectly
from the symmetry algebra in Section 3)
from a covariant action. 
This represents a consistency check  on the two 
different  methods used in Section 3 and in the present Section.

\subsection{Equations of motion}

The
equations of motion corresponding to  the phase space 
superstring Lagrangian \rf{lag6} are
\begin{eqnarray}
\label{eqmot1}
\dot{\phi} &= &\frac{e^{2\phi}}{p^+}\Pi\,,
\qquad
\dot{\Pi}
= \frac{1}{p^+}\partial_\sigma (e^{2\phi}\ph')
+2\PP^-\,,
\\
\dot{u}^M\!
&=&\frac{e^{2\phi}}{p^+}\PP^M
-{\rm i}e^{2\phi}\eta_i(\sigma^{MN})^i{}_j\eta^j u^N \ , 
\\
\dot{\PP}^M\!
& =& -\frac{e^{2\phi}}{p^+}u^M \PP^N\PP^N
+\frac{1}{p^+}v^{MN}\partial_\sigma(e^{2\phi}\u'^N)
-{\rm i}e^{2\phi}\eta_i(\sigma^{MN})^i{}_j\eta^j \PP^N
\\
&+& 
e^{2\phi}v^{MN}\eta^i\rho_{ij}^N\th'^j
+e^{2\phi}v^{MN}\eta_i(\rho^N)^{ij}\th'_j
\\
\dot{\theta^i}
&=&
\frac{\rm i}{p^+}\partial_\sigma(e^{2\phi}\eta_j y^{ji})
\ , \ \ \ \ \ \ 
\dot{\theta_i}
=
\frac{\rm i}{p^+}\partial_\sigma(e^{2\phi}\eta^j y_{ji}) \ , 
\\
\label{eqmot11}
\dot{\eta}^i
&=&
e^{2\phi}
\Bigl[{\rm i}\eta^2\eta^i
-\frac{2\rm i}{p^+}(l\eta)^i
+\frac{\rm i}{p^+}y^{ij}\th'_j \Bigr]\ , 
\quad
\dot{\eta}_i
=e^{2\phi}
\Bigl[-{\rm i}\eta^2\eta_i
+\frac{2\rm i}{p^+}(\eta l)_i
+\frac{\rm i}{p^+}y_{ij}\th'^j\Bigr],\hspace{0.8cm}
\end{eqnarray}
where $v^{MN}$ is given by (\ref{vmn})
and, as  previously,  do not distinguish between  the upper and lower
``$S^3$"  
indices $M,N$, i.e.  use  the convention $\PP_M = \PP^M$.
These equations can be written in the  Hamiltonian form.  
Introducing  
the notation $\cal X$
for the phase space variables  $(\Pi,\phi,\PP^M,u^M,
\theta^i,\theta_i,\eta^i,\eta_i)$, the Hamiltonian
equations are 
\be\label{hamfor}
\dot{\cal X} = [{\cal X},\PP^-] \ , 
\ee
where  the phase space variables
satisfy the (classical) Poisson-Dirac brackets
\be\label{qcom1}
[\ \Pi(\sigma), \phi(\sigma')\ ] =\delta(\sigma,\sigma')\,,
\ee
\be\label{npcom}
[\PP^M(\sigma),u^N(\sigma')]
=v^{MN}\delta(\sigma,\sigma')\,,
\qquad
[\PP^M(\sigma),\PP^N(\sigma')]
=(u^M \PP^N - u^N \PP^M ) \delta(\sigma,\sigma')\,,
\ee
\be\label{ttcom}
\{\theta_i(\sigma), \theta^j(\sigma')\} 
= \frac{\rm i}{p^+}\delta_i^j\delta(\sigma,\sigma')\ , 
\qquad
\{\eta_i(\sigma), \eta^j(\sigma')\}  
=\frac{\rm i}{p^+}\delta_i^j\delta(\sigma,\sigma') \ , 
\ee
\be\label{qcom6}
[x_0^-,\theta^i]=\frac{1}{2p^+}\theta^i,
\quad
[x_0^-,\theta_i]=\frac{1}{2p^+}\theta_i\,,
\quad
[x_0^-,\eta^i]=\frac{1}{2p^+}\eta^i,
\quad
[x_0^-,\eta_i]=\frac{1}{2p^+}\eta_i\,.
\ee
$x_0^-$ is the  zero mode of $x^-$ so that   $[p^+,x_0^-]=1$.
All  the remaining brackets are equal
to zero. 
The structure of  (\ref{npcom}) reflects the fact that in the Hamiltonian
formulation the condition   $u^Mu^M=1$ should be supplemented by 
the constraint
\be\label{seccon}
u^M\PP^M=0\,.
\ee 
These  are second class constraints,  and the Dirac
procedure leads then to the classical Poisson-Dirac brackets (\ref{npcom}).
To derive     (\ref{ttcom}),(\ref{qcom6})  one is to take into account
 that  the Lagrangian 
(\ref{lag6}) has the following second class constraints

\be
 p_{\theta^i}+\frac{\rm i}{2}p^+\theta_i=0\,,
\qquad
 p_{\theta_i}+\frac{\rm i}{2}p^+\theta^i=0 \ , 
\ee
where $p_{\theta^i}$, $p_{\theta_i}$ are the canonical momenta of fermionic
coordinates. The same constraints are found for the fermionic coordinates
$\eta^i$,
$\eta_i$.  Starting with the  Poisson brackets

\be
\{ p_{\theta^i},\theta^j\}_{_{P.B.}}=\delta_i^j\,,
\qquad
\{ p_{\theta_i},\theta_j\}_{_{P.B.}}=\delta^i_j\,,
\qquad
[p^+,x_0^-]_{_{P.B.}}=1\,,
\ee
one  gets then the 
 Poisson-Dirac brackets given in (\ref{ttcom}),(\ref{qcom6}).

\newsection{Noether charges as generators of the superalgebra
$psu(1,1|2)\oplus\widetilde{psu}(1,1|2)$ }

The Noether charges play an  important role in 
the analysis of the symmetries of
dynamical systems. The choice of the light-cone gauge spoils 
manifest global
symmetries,  and  in order to demonstrate that these  global
invariances 
are still
present   one  needs to  find  the  Noether charges
which generate them.\foot{In what follows ``currents'' and
``charges''
will mean both bosonic and fermionic ones, 
i.e.  will include supercurrents  and supercharges.}
These charges determine  
  the structure of   superstring   field theory 
 in the \lc gauge \ci{GSB}.
The first step in  the  construction  of superstring field 
theory is to find  a  free (quadratic) superfield 
representation of the generators of the 
$psu(1,1|2)\oplus\widetilde{psu}(1,1|2)$
superalgebra. The   charges we  obtain  below
can be used  to  obtain  (after quantization) 
these free superstring field  charges.

The Noether charges for  a superparticle in $AdS_3\times S^3$ were 
found in Section 3. These charges are helpful in
establishing a correspondence
between the bulk fields of type  IIB supergravity and  
the chiral primary operators of the boundary  theory
in a manifestly  supersymmetric  way. 
Superstring Noether charges  should   thus be important 
 for the  study
of the AdS/CFT correspondence at the full string-theory level.
Our discussion below will be an 
adaptation to the $AdS_3\times S^3$ case 
of the results for the currents in the 
$AdS_5\times S^5$ case given in  \ci{MTT}.

\subsection{Currents for  $\kappa$-symmetry 
light-cone  gauge  fixed\\   superstring action}

As usual,   symmetry  generating charges   can be obtained from
conserved currents. 
Since currents  themselves 
may be helpful in  some  applications, we shall  first
derive  them starting with the  
$\kappa$-symmetry gauge fixed  Lagrangian in the form given in 
(\ref{actkin6}),(\ref{actwz6}) and   using 
the 
standard Noether method based on the
localization of the parameters of
the  associated global transformations.
 Let $\epsilon$ be a parameter of some global transformation 
which leaves the action  invariant. Replacing it by a 
function of worldsheet coordinates $\tau,\sigma$,   the variation of
 the 
action takes the form
\be\label{dS}
\delta S =\int d^2\sigma\  {\cal G}^\vm \partial_\vm \epsilon\, , 
\ee
where ${\cal G}^\vm$  is the  corresponding current.
Making use of this formula, we shall  find below those currents
which
are related to  symmetries that do not 
involve compensating $\kappa$-symmetry transformation. 
The remaining currents will  be found  in the next subsection 
starting  from  the 
action   \rf{lag6}  where   both the   $\kappa$-symmetry and 
the bosonic  light-cone gauges are fixed.

Let us  start with  the translation invariance 
$\delta x^a = \epsilon^a$.
Applying  (\ref{dS}) to the Lagrangian 
(\ref{actkin6}),(\ref{actwz6}) gives the translation currents
\begin{eqnarray}
\label{con1}
{\cal P}^{+\vm}
&=&-\sqrt{g}g^{\vm\vn}e^{2\phi}\partial_\vn x^+\ , 
\\
{\cal P}^{-\vm}
&=&-\sqrt{g}g^{\vm\vn}\Bigl(e^{2\phi}\partial_\vn x^-
+F^M D_\vn u^M\Bigr) \nonumber
\\
&-&
\frac{\rm i}{2}\sqrt{g}g^{\vm\vn}e^{2\phi}\Bigr(
\theta^i\partial_\vn\theta_i
+\theta_i\partial_\vn\theta^i
+\eta^i\partial_\vn\eta_i
+\eta^i\partial_\vn\eta^i+2{\rm i}e^{2\phi}\partial_\vn
x^+(\eta^2)^2\Bigr)
\nonumber
\\
&+&
\epsilon^{\vm\vn}e^{2\phi}(\eta^i y_{ij}\partial_\vn\theta^j
+\eta_i y^{ij}\partial_\vn\theta_j)\,,
\quad
F^M \equiv {\rm i}\eta_i(\sigma^{MN})^i{}_j \eta^j e^{2\phi}u^N\,.
\label{con4}
\hspace{1cm}
\end{eqnarray}
Some of the  remaining bosonic currents can be expressed 
in terms of  supercurrents.
The invariance with respect to the  super-transformations
\be
\delta\theta^i = \epsilon^i\,,
\qquad
\delta\theta_i  = \epsilon_i\,,
\qquad
\delta x^- 
= -\frac{\rm i}{2}\epsilon^i\theta_i 
- \frac{\rm i}{2}\epsilon_i\theta^i \ , 
\ee
gives the  supercurrents
\be\label{con8}
{\cal Q}^{+i\vm}
=-\sqrt{g}g^{\vm\vn}e^{2\phi}\theta^i\partial_\vn x^+
+{\rm i}\epsilon^{\vm\vn}e^{2\phi}\eta_j y^{ji}\partial_\vn x^+ \ , 
\ee
\be\label{con9}
{\cal Q}_i^{+\vm}
=-\sqrt{g}g^{\vm\vn}e^{2\phi}\theta_i\partial_\vn x^+
+{\rm i}\epsilon^{\vm\vn}e^{2\phi}\eta^j y_{ji}\partial_\vn x^+ \ . 
\ee
The invariance of the action (\ref{actkin6}),(\ref{actwz6}) 
with respect to the rotation of (super) coordinates 
in the  $(x^+,x^-)$ plane
\be
\delta x^\pm =e^{\pm \epsilon} x^\pm\,,
\qquad
\delta(\theta^i,\theta_i\,,\eta^i,\eta_i)
=e^{-\epsilon/2}(\theta^i,\theta_i\,,\eta^i,\eta_i) \ , 
\ee
leads to 
\be\label{con11}
{\cal J}^{+-\vm}
=x^+{\cal P}^{-\vm}- x^-{\cal P}^{+\vm}
+\frac{\rm i}{2}\theta^i{\cal Q}_i^{+\vm}
+\frac{\rm i}{2}\theta_i{\cal Q}^{+i\vm} \ ,  
\ee
while the  invariance with respect to  the dilatations
\be
\delta x^a = e^\epsilon x^a\,,
\qquad
\delta\phi =-\epsilon\,,
\qquad
\delta(\theta^i,\theta_i\,,\eta^i,\eta_i)
=e^{\epsilon/2}(\theta^i,\theta_i\,,\eta^i,\eta_i) \ , 
\ee
implies conservation of   the dilatation current
\be\label{con15}
{\cal D}^\vm
=x^a{\cal P}^{a\vm}+\sqrt{g}g^{\vm\vn}\partial_\vn \phi
-\frac{\rm i}{2}\theta^i{\cal Q}_i^{+\vm}
-\frac{\rm i}{2}\theta_i{\cal Q}^{+i\vm} \ . 
\ee
The invariances with respect to  the $SU(2)$ 
($ \epsilon^i{}_i=0$)  and 
 $\widetilde{SU}(2)$ rotations ($ \tilde{\epsilon}^i{}_i=0$) 
\be   
\delta y^{ij} = \epsilon^i{}_l y^{lj}\,,
\quad{\rm i.e.}
\quad  
\delta u^M =-\frac{1}{2}\epsilon^i{}_j(\sigma^{MN})^j{}_i u^N,
\qquad
\delta \eta^i =\epsilon^i{}_j\eta^j\,,
\quad 
\delta \eta_i =-\eta_j \epsilon^j{}_i\,,
\ee
\be   
\delta y^{ij} = \tilde{\epsilon}^j{}_l y^{il}, 
\quad
{\rm i.e.}  
\quad \delta u^M =-\frac{1}{2}\tilde{\epsilon}^i{}_j
(\bar{\sigma}^{MN})^j{}_i u^N,
\qquad
\delta \theta^i =\tilde{\epsilon}^i{}_j\theta^j\,,
\quad \ \ \  
\delta \theta_i =-\theta_j \tilde{\epsilon}^j{}_i\,,
\ee
give the following $SU(2)$ and $\widetilde{SU}(2)$ currents, 
respectively, 
\be\label{con16}
{\cal J}^i{}_j^{ \ \vm  } 
=-{\rm i}\sqrt{g}g^{\vm\vn}
\frac{1}{2}(\sigma^{MN})^i{}_ju^MD_\vn u^N
+(\eta^i\eta_j-\frac{1}{2}\delta^i_j\eta^2)\PP^{+\vm}\,.
\ee
\begin{eqnarray}\label{con17}
{}\hspace{-1cm}\widetilde{{\cal J}}^i{}_j^{ \ \vm  } 
=&-&{\rm i}\sqrt{g}g^{\vm\vn}
\frac{1}{2}(\bar{\sigma}^{MN})^i{}_ju^MD_\vn u^N
+(\theta^i\theta_j-\frac{1}{2}\delta^i_j\theta^2)\PP^{+\vm}
\nonumber\\
&-&{\rm i}
\epsilon^{\vm\vn}e^{2\phi}\partial_\vn x^+(\eta^l y_{lj} \theta^i
-\frac{1}{2}\delta^i_j \eta^k y_{kl}\theta^l)
+{\rm i}\epsilon^{\vm\vn}e^{2\phi}\partial_\vn x^+(
\eta_l y^{li}\theta_j
-\frac{1}{2}\delta^i_j \eta_k y^{kl}\theta_l)\,.
\end{eqnarray}

\subsection{Charges for  bosonic and $\kappa$-symmetry 
  light-cone gauge fixed \\
superstring action}

In  the previous Section we have listed the  (super)currents starting
with the  $\kappa$-symmetry  \lc gauge fixed  action given in
(\ref{actkin6}),(\ref{actwz6}).  They  can be used  to
find currents for  the  action where  both the fermionic
$\kappa$-symmetry and the  bosonic  reparametrization symmetry are
fixed by the  \lc type gauges   (\ref{lag6}).  To find the
components of currents (${\cal G}^0$)
 in the world-sheet time direction
one needs to use the relations  
(\ref{can1})--(\ref{can6})  for  the canonical momenta and to 
  insert the  light-cone gauge conditions  (\ref{lcg}) and
  (\ref{h00}) 
into   the expressions for the currents. 
The  charges are then 
$G =\int d\sigma\  {\cal G}^0$.

Let us start with the kinematical generators (charges) (\ref{kingen}).
The results for the currents imply the following representations
\be\la{kin1}
P^+ = p^+\,,
\qquad
Q^{+i} = \int p^+ \theta^i\,,
\qquad
Q_i^+ = \int  p^+\theta_i\ . 
\ee
Note that these charges depend only on the zero modes of string
coordinates
(the integrands are ${\cal G}^0$ parts
of the corresponding currents in world-sheet time direction:
${\cal Q}^{+i 0}$,
${\cal Q}_i^{+ 0}$
and
${\cal P}^{+0}=p^+$).
The remaining kinematical charges depend on non-zero string modes
\begin{eqnarray}
&&
J^{+-} = \int x^+\PP^- -x^- p^+ \ , 
\qquad\ \ 
D = 
\int 
x^+\PP^-  + x^- p^+  - \Pi\,,
\\
&&
J^i{}_j = \int l^i{}_j + p^+\eta^i\eta_j
-\frac{1}{2}\delta_j^i p^+\eta^2\ , 
\qquad\widetilde{J}^i{}_j 
= \int \tilde{l}^i{}_j +p^+\theta^i \theta_j
-\frac{1}{2}\delta_j^i p^+\theta^2\ , 
\end{eqnarray}
where $l^i{}_j$ is given by (\ref{lcla}) 
and $\widetilde{l}^i{}_j 
=\frac{\rm i}{2}(\bar{\sigma}^{MN})^i{}_j u^M \PP^N$.
The derivation of the  remaining charges follows the procedure described
in  Appendix D of \cite{MTT}. The conformal (super)charges are given by
(\ref{kpxp}),(\ref{spxp}) where
\be\label{den1}
S_i^{+}|_{x^+=0}
=\int \frac{1}{\sqrt{2}}e^{-\phi} p^+\eta_i\,,
\qquad
S^{+i}|_{x^+=0}
= 
\int \frac{1}{\sqrt{2}}e^{-\phi} p^+\eta^i\,,
\ee
\be
K^+|_{x^+=0} =
\int
-\frac{1}{2}e^{-2\phi}p^+\,.
\ee
The dynamical  Poincar\'e 
charges $Q^{-i}$, $Q^-_i$ and the conformal charges $S^{-i}$, $S^-_i$
are 
\begin{eqnarray}
\label{q1}
&&
Q_i^- =\int \frac{e^\phi}{\sqrt{2}}
\Bigl({\rm i}\eta_i\Pi
-p^+\eta^2\eta_i + 2 \eta_j l^j{}_i
+y_{ij}\th'^j \Bigr) \ , 
\\
\label{q2}&&
Q^{-i} =\int \frac{e^\phi}{\sqrt{2}}
\Bigl(-{\rm i}\eta^i\Pi
-p^+\eta^2\eta^i + 2 l^i{}_j\eta^j
-y^{ij}\th'_j\Bigr)\,.
\end{eqnarray}
\be\label{sden1}
S^{-i} = \int \theta^i S -\tilde{l}^i{}_j\theta^j
+\frac{1}{2}e^{-\phi}\partial_\sigma (e^\phi\eta_j)\ y^{ji}
\ee
\be\label{sden2}
S^-_i = \int \theta_i \bar{S} -\theta_j \tilde{l}^j{}_i
-\frac{1}{2}e^{-\phi}\partial_\sigma (e^\phi\eta^j)\ y_{ji}
\ee
\be
S={\rm i}x^-p^+ - \frac{\rm i}{2}\Pi +\frac{1}{2}p^+ \theta^2\ , \ \ 
\qquad
\bar{S}
=-{\rm i}x^-p^+ + \frac{\rm i}{2}\Pi +\frac{1}{2}p^+ \theta^2\ . 
\ee
Note that the $G|_{x^+=0}$ parts of the kinematical charges (\ref{kingen}) 
can be 
obtained from the  superparticle  ones simply by replacing the 
particle coordinates by the string ones.
The remaining dynamical generator $K^-$ can be found  by using the 
expressions found above and  applying  the  commutation 
relations of $psu(1,1|2)\oplus \widetilde{psu}(1,1|2)$ superalgebra.

Our classical charges are normalized so that after the  
quantization, i.e. the 
replacement  of the  classical 
Poisson-Dirac brackets (\ref{qcom1})-(\ref{qcom6}) by 
quantum (anti)commutators 
\be
[\ ,\ ]_{_{P.B}} \rightarrow {\rm i}[\ ,\ ]\,,
\qquad
\{ \ ,\ \}_{_{P.B}} \rightarrow {\rm i}\{\ ,\ \}\,,
\ee
redefinitions
$J^{+-} \rightarrow -{\rm i}J^{+-}$,  
$D \rightarrow -{\rm i}D$, 
$K^\pm \rightarrow -K^\pm$,
and appropriate   operator ordering the charges satisfy
the commutation relations of  $psu(1,1|2)$ $\oplus$ 
$\widetilde{psu}(1,1|2)$ superalgebra given
in (\ref{lccom1})-(\ref{lccom11}).

\section*{Acknowledgments}

The  work  of R.R.M. and A.A.T.  was  supported  by
the DOE grant DE-FG02-91ER-40690 
and the INTAS project 991590.
 R.R.M. is also supported by the RFBR Grant No.99-02-16207.
A.A.T. would like to acknowledge also the support of the EC TMR 
grant ERBFMRX-CT96-0045 and the PPARC SPG grant  PPA/G/S/1998/00613.


\setcounter{section}{0}
\setcounter{subsection}{0}


\appendix{Notation and basic definitions}
\label{not}

In the main part of the paper we 
 use the following  conventions for the indices:
\begin{eqnarray*}
a, b  =0, 1 & &\qquad
\hbox{boundary Minkowski space indices}
\\
{\cal A}, {\cal B}, {\cal C}=1,2,3 & &\qquad
\hbox{$S^3$  coordinate space indices}
\\
\apr,\bpr,\cpr=1,2,3 && \qquad  so(3)\  \hbox{ vector
indices ($S^3$ tangent space indices) }
\\
M,N,K,L =1,\ldots, 4
&& \qquad so(4)\  \hbox{vector indices}
\\
i,j,k,n =1,2 && \qquad su(2) \hbox{ and } \widetilde{su}(2)\  
\hbox{vector indices}
\\
\mu,\nu = 0,1 
&&
\qquad
\hbox{world-sheet coordinate indices}
\end{eqnarray*}
We decompose $x^a$ into the light-cone coordinates
$x^a= (x^+,x^-)$ where
$x^\pm\equiv \frac{1}{\sqrt{2}}(x^1\pm x^0)$.
We suppress the flat space metric tensor $\eta_{ab}=(-,+)$ 
in scalar products, i.e. 
$A^a B^a\equiv \eta_{ab}A^a B^b $.
The $SO(1,1)$ vector $A^a$  is decomposed as
$A^a = (A^+,A^-)$  so that the scalar product is 
$A^a B^a = A^+B^- + A^- B^+$.
The  derivatives with respect to the  world-sheet coordinates
$(\tau,\sigma)$ are 
\be
\dot{x} \equiv \partial_\tau x\,,\ \ \  
\qquad
\x' \equiv \partial_\sigma x
\ee
The  world-sheet 
Levi-Civita $\epsilon^{\vm\vn}$   is defined with   $\epsilon^{01}=1$.

The four  matrices   $(\sigma^M)^i{}_j$,
$({\bar\sigma^M})^i{}_j$ are off-diagonal blocks of  
the $SO(4)$   Dirac   matrices $\gamma^M$
in the  chiral (Weyl) representation, i.e. 
\be\label{usgam}
\gamma^M
=\left(\begin{array}{cc}
 0   & \sigma^M
 \\
 \bar{\sigma}^M & 0
 \end{array}
 \right)\,, 
 \quad\ \ 
(\sigma^M)^i{}_k(\bar{\sigma}^N)^k{}_j 
+(\sigma^N)^i{}_k(\bar{\sigma}^M)^k{}_j 
 =2\delta^{MN}\delta_j^i\,,
\ee
\be
 C_{ik}^\prime(\sigma^M)^k{}_j 
 = C_{jk}^\prime(\bar{\sigma}^M)^k{}_i\,,
 \quad (\sigma^M)^i{}^*_j\equiv  (\bar{\sigma}^M)^j{}_i \ , 
 \ee
 where $C'$ is a charge conjugation matrix.
\  $\sigma^{MN}$, $\bar{\sigma}^{MN}$  are defined by 
\be
(\sigma^{MN})^i{}_j \equiv \frac{1}{2}(\sigma^M)^i{}_k
(\bar{\sigma}^N)^k{}_j
-(M\leftrightarrow N)\,, 
\quad
(\bar{\sigma}^{MN})^i{}_j \equiv \frac{1}{2}(\bar{\sigma}^M)^i{}_k
(\sigma^N)^k{}_j
-(M\leftrightarrow N)\,.  
\la{rrr}
\ee
We use the following explicit form of  $\sigma^M$, 
$\bar{\sigma}^M$ and $C_{ij}^\prime$ 
\be
\sigma^M = (\sigma^\apr, -{\rm i}I)\,,
\qquad
\bar{\sigma}^M = (\sigma^\apr, {\rm i}I)\,,
\qquad
C_{ij}^\prime = c\epsilon_{ij}\,,
\quad
|c|=1\,.
\ee
We also use the matrices  $\rho^M$  defined by
\be
\rho^M_{ij} \equiv C_{ik}^\prime (\sigma^M)^k{}_j\,,
\qquad
(\rho^M)^{ij} \equiv -(\sigma^M)^i{}_k (C^{\prime -1})^{kj}\,,
\quad
(\rho_{ij}^M)^* = -(\rho^M)^{ij}
\ee
We assume  the following Hermitian conjugation rule for the 
fermionic coordinates and the notation for their squares 
\be
\theta_i^\dagger =\theta^i\,,
\qquad
\eta_i^\dagger = \eta^i\ , 
\qquad
\theta^2 \equiv \theta^i\theta_i\,,
\qquad \ \ \  
\eta^2
 \equiv \eta^i\eta_i\,.
\ee
The generators of the 
$su(2)$ and $\widetilde{su}(2)$ subalgebras of $so(4)$ 
\be
[J^{MN},J^{KL}]=\delta^{NK}J^{ML}+ \hbox{ 3 terms}
\ee
are defined  by 
\be
J^i{}_j =\frac{1}{4}(\sigma^{MN})^i{}_j J^{MN}\,,
\qquad\widetilde{J}^i{}_j =\frac{1}{4}(\bar{\sigma}^{MN})^i{}_j J^{MN}\,.
\ee
The translation operator $J^{4\apr}$ on $S^3$ is 
\be
J^{4\apr} = \frac{\rm i}{2}(\sigma^\apr)^j{}_i (J^i{}_j - \tilde{J}^i{}_j),.
\ee
The coset representative of $S^3$ defined 
by $g_y\equiv \exp(y^\apr J^{4\apr})$
takes then the form given in (\ref{gy}) below.
In terms of these coordinates,  
the 3-sphere  interval, metric tensor
and  vielbein are given by
\be
ds_{S^3}^2=d|y|^2+ {\rm sin}^2|y| ds_{S^2}^2 \ , \ \ \ \  
\    ds_{S^2}^2 = dn^\sca dn^\sca \   ,  \ \ \ n^\sca n^\sca =1 \ ,  
\ee 
\begin{equation}\label{b1}
G_{\sca\scb}= e_\sca^\apr e_\scb^\apr\,,   \ \
\qquad
e_\sca^\apr=\frac{\sin |y|}{|y|}(\delta_\sca^\apr-n_\sca n^\apr)
+n_\sca n^\apr    \ , 
\end{equation}
\be\label{b2}
G_{\sca\scb}= 
\frac{\sin^2|y|}{|y|^2}(\delta_{\sca\scb}-n_\sca n_\scb)
+n_\sca n_\scb \ ,    
\qquad
n^\sca \equiv \frac{y^\sca}{|y|}\,,
\qquad
|y| =\sqrt{y^\apr y^\apr}\,.
\ee
We use the convention
 $y^\sca = \delta_\apr^\sca y^\apr$ and the same
for $n^\apr$. The $S^3$  Killing vectors  $V^\apr$ and
$V^{\apr\bpr}$ corresponding to the 3  translations and 3
$SO(3)$ rotations are  
\be
\label{b4}
V^\apr
=\Bigl[|y|\cot |y| (\delta^{\apr\sca}-n^\apr n^\sca)
+n^\apr n^\sca\Bigr]\partial_{y^\sca}  \ ,
\quad
V^{\apr\bpr}=y^\apr \partial_{y^\bpr}-y^\bpr \partial_{y^\apr}  \ .
\ee
They can be collected into the  $SU(2)$ combination 
\begin{equation}\label{b5}
(V^\sca)^i{}_j\partial_{y^\sca}
=\frac{1}{4}(\sigma^{\apr\bpr})^i{}_jV^{\apr\bpr}
+\frac{\rm i}{2}(\sigma^\apr)^i{}_j V^\apr      \ .
\end{equation}
These relations and (\ref{uapr}) imply the following  relations
\begin{equation}\label{2rel}
G_{\sca\scb} (\eta V^\sca \eta) ( \eta V^\scb\eta)
=(\eta R^M\eta )^2\ , \ \ \ 
\qquad
G_{\sca\scb} (\eta V^\sca \eta) dy^\scb = 
(\eta R^M \eta) du^M \ , 
\end{equation}
where $R^M$ is defined by (\ref{rm}), 
which have been used to transform (\ref{actkin3}) to (\ref{actkin6}).


\appendix{Light-cone basis  
of  $psu(1,1|2)\oplus \widetilde{psu}(1,1|2)$}


Here we explain the relation   between 
the $su(1,1)\oplus su(2)$ covariant and 
light-cone
bases of the $psu(1,1|2)$ algebra and define the light-cone basis of
$psu(1,1|2)\oplus \widetilde{psu}(1,1|2)$. We find it convenient to introduce
intermediate basis defined by 
\be
m^{+-} \equiv -m^1{}_1\,,
\qquad
m^{+ 1}\equiv \frac{1}{\sqrt{2}}m^2{}_1\,,
\qquad
m^{- 1}\equiv -\frac{1}{\sqrt{2}}m^1{}_2\,,
\ee
\be
q^{+ i} \equiv  - q_1^i\,,
\qquad
q^{- i} \equiv q_2^i\,,
\qquad
q^+_i \equiv q^2_i\,,
\qquad
q^-_i \equiv q^1_i\,.
\ee
In this basis the Hermitean rules for supercharges take conventional 
form 
$(q^{+ i})^\dagger = q^+_i$, 
$(q^{- i})^\dagger = q^-_i$, and  for the
(anti)commutators one has
\be
[m^{+-},m^{\pm 1}]= \pm m^{\pm 1}\,,
\qquad
[m^{+1},m^{-1}]= -m^{+-}\,,
\qquad
[m^{+-},q_i^\pm]=\pm \frac{1}{2}q^\pm_i\,,
\ee
\be
\{q^{\pm i},q^\pm_j\} = -a \sqrt{2}\delta^i_j m^{\pm 1}\,,
\qquad
\{q^{\pm i},q_j^\mp\} =a(\delta_j^i m^{+-} \mp m^i{}_j)\,,
\ee
\be
[q^{- i},m^{+ 1}] = -\frac{1}{\sqrt{2}}q^{+ i}\,,
\qquad
[q^{+ i},m^{- 1}] = \frac{1}{\sqrt{2}}q^{- i}\ . 
\ee
The light-cone basis of $psu(1,1|2)\oplus \widetilde{psu}(1,1|2)$ superalgebra
is defined by 
\be
P^+ = \sqrt{2}\tilde{m}^{+1}\,,
\qquad
P^- = \sqrt{2}m^{-1}\,,
\qquad
K^+ = \sqrt{2}m^{+1}\,,
\qquad
K^- = \sqrt{2}\tilde{m}^{-1}\ , 
\ee
\be
J^{+-} = m^{+-} + \tilde{m}^{+-}\,,
\qquad
D = m^{+-} - \tilde{m}^{+-}\,,
\qquad
J^i{}_j = m^i{}_j\,,
\qquad\widetilde{J}^i{}_j = \tilde{m}^i{}_j\,,
\ee
\be
Q^{+i} = \tilde{q}^{+i}\,,
\qquad
Q^+_i=\tilde{q}^+_i\,,
\qquad
Q^{-i} = q^{-i}\,,
\qquad
Q^-_i = q^-_i\ , 
\ee
\be
S^{-i} = \tilde{a}\tilde{q}^{-i}\,,
\qquad
S^-_i = \tilde{a}^* \tilde{q}^-_i\,,
\qquad
S^{+i} = a q^{+i}\,,
\qquad
S^+_i =a^* q^+_i\,.
\ee
The constants $a, \tilde a$ are 
chosen to be  $a=-{\rm i}$, $\tilde{a}={\rm i}$.
Then the  commutation relations are
\begin{eqnarray}
&[P^\pm, K^\mp] = D \mp J^{+-}\ , 
&
\nonumber
\\
&
[D,P^\pm] =-P^\pm\,,
\quad
[D,K^\pm] =K^\pm\,,
\quad
[J^{+-},P^\pm]=\pm P^\pm\,,
\quad
[J^{+-},K^\pm]=\pm K^\pm\,,
&
\nonumber
\\
&
[D,Q_i^\pm] =-\frac{1}{2}Q_i^\pm\,,
\quad
[D,S_i^\pm] = \frac{1}{2}S_i^\pm\,,
\quad
[J^{+-},Q_i^\pm]=\pm \frac{1}{2}Q_i^\pm\,,
\quad
[J^{+-},S_i^\pm]=\pm\frac{1}{2} S_i^\pm\,,
&\nonumber
\\
&{}\!\![S^\pm _i,P^\mp]={\rm i}Q^\mp_i\,,
\quad
\![Q^{\pm i},K^\mp ]=-{\rm i}S^{\mp i},
\quad
\!\{Q^{\pm i},Q^\pm _j\}=\mp {\rm i}P^\pm \delta_j^i,
\quad
\!\!\{S^{\pm i},S^\pm _j\}=\pm{\rm i}K^\pm \delta_j^i
&
\nonumber
\\
\label{lccom1ap}
&
\{Q^{+i},S^-_j\}=\frac{1}{2}(J^{+-} - D)\delta_j^i
-\tilde{J}^i{}_j\,,
\qquad
\{Q^{-i},S^+_j\}=\frac{1}{2}(J^{+-} + D)\delta_j^i
+J^i{}_j\ . 
&
\end{eqnarray}
The supercharges $Q^-_i$, $Q^{-i}$, $S^{+i}$, $S^+_i$
transform in the  fundamental representation 
of $su(2)$ i.e. they are rotated 
only by $J^i{}_j$ and satisfy  (\ref{lccom5}).
The remaining supercharges 
$Q^{+i}$, $Q^+_i$, $S^{-i}$, $S^-_i$
transform in fundamental representation 
of $\widetilde{su}(2)$ i.e. 
they are rotated only by $\tilde{J}^i{}_j$ 
and satisfy  (\ref{lccom6}).
All the generators  except $K^a$ and $P^a$ satisfy the
Hermiteant conjugation rules  in (\ref{herrul}), while  
$K^a$ and $P^a$
are taken to be anti-Hermitean:
$(P^\pm)^\dagger= - P^\pm$,
$(K^\pm)^\dagger= - K^\pm$.
The light-cone basis for generators 
described above is used  in the  calculation of the 
Cartan forms in Appendix E. 
Making the substitutions  $P^\pm \rightarrow 
{\rm i}P^\pm$, $K^\pm \rightarrow -{\rm i}K^\pm$ we obtain the 
basis  used in Sections 2-4.


\appendix{Derivation of supercharges}


Here we would like to demonstrate how the knowledge  of kinematical charges 
and commutation relations of superalgebra allows
one to get dynamical charges systematically. Consider, for example, 
 the
dynamical
supercharges $Q^{-i}$ whose most general form is
\be
Q^{-i}
= Q^{-i}(p^+,\partial_{p^+},z,\partial_z,
\theta,\lambda, \eta,\vartheta)\,,
\ee
where a dependence on $S^3$ is not shown explicitly.
{}From  $[P^+,Q^{-i}]=0$ we get
\be
Q^{-i}={}^{^{(1)}}\!Q^{-i}(p^+,z,\partial_z,
\theta,\lambda, \eta,\vartheta)\,,
\ee
i.e. we learn that $Q^{-i}$ does not depend on $\partial_{p^+}$. From 
$\{Q^{-i},Q^{+j}\}=0$ we get
\be
{}^{^{(1)}}\!Q^{-i}(p^+,z,\partial_z, \theta,\lambda, \eta,\vartheta)
={}^{^{(2)}}\!Q^{-i}(p^+,z,\partial_z,\theta,\eta,\vartheta)\,,
\ee
i.e.  $Q^{-i}$ does not depend on $\lambda$.
The anticommutator 
$\{Q^{-i},Q_j^+\}=0$ tell us that  
$Q^{-i}$ does not depend on $\theta$, i.e.
\be
{}^{^{(2)}}\!Q^{-i}(p^+,z,\partial_z,\theta,\eta,\vartheta)
={}^{^{(3)}}\!Q^{-i}(p^+,z,\partial_z,\eta,\vartheta)\,.
\ee
{}From  $[Q^{-i},K^+]=S^{+i}$
we get
\be
{}^{^{(3)}}\!Q^{-i}=\frac{1}{\sqrt{2}}\eta^i\partial_z+
{}^{^{(4)}}\!Q^{-i}(p^+,z,\eta,\vartheta)\,,
\ee
i.e.
\be
Q^{-i}
=\frac{1}{\sqrt{2}}\eta^i\partial_z+
{}^{^{(4)}}\!Q^{-i}(p^+,z,\eta,\vartheta)\ , 
\ee
and from  $\{Q^{-i},S^{+j}\}=0$ we get
\be
{}^{^{(4)}}\!Q^{-i}
=-\frac{1}{\sqrt{2}z}\eta^i(\eta\vartheta)
+{}^{^{(5)}}\!Q^{-i}(p^+,z,\eta)
\ee
i.e. 
\be\label{c8}
Q^{-i}=\frac{1}{\sqrt{2}}\eta^i\partial_z
-\frac{1}{\sqrt{2}z}\eta^i(\eta\vartheta)
+{}^{^{(5)}}\!Q^{-i}(p^+,z,\eta)\ . 
\ee
The second anticommutator in (\ref{lccom4}) gives
\be\label{c9}
{}^{^{(5)}}\!Q^{-i}(p^+,z,\eta)
=\frac{1}{2\sqrt{2}z}\eta^i
+\frac{2}{\sqrt{2}z} (l\eta)^i
+{}^{^{(6)}}\!Q^{-i}(p^+,z)\ , 
\ee
and the $su(2)$ covariance implies ${}^{^{(6)}}\!Q^{-i}(p^+,z)=0$.
Taking this into account and plugging (\ref{c9}) into 
 (\ref{c8}) we get
$Q^-_i$ given by (\ref{q2den}).
Using the Hermitean conjugation rule 
$(Q^{-i})^\dagger=Q^-_i$ we get
the expression for $Q_i^-$  in (\ref{q1den}).
The  anticommutator $\{Q_i^-,Q^{-j}\}= -P^-\delta_i^j$ 
determines $P^-$, i.e.  the Hamiltonian.
The remaining dynamical generators
 $K^-$, $S^{-i}$, $S_i^-$ can be obtained in a similar 
way.


\appendix{Eigenvectors of $AdS$ mass operator }


Here we would like to 
explain the  procedure of finding  the 
eigenvectors of the AdS mass  operator $A$ or the operator 
$X$ in \rf{adef}
in Section 4.1.  Since  the 
superfield
$\Phi_{k,\sigma}$ in (4.11) diagonalizes the  operators $l^2$ and
$\eta\vartheta$  we have to diagonalize the 
operator $\vartheta l \eta\equiv  \vartheta_i l^i_{\ j}  \eta^j  $, 
i.e. to  find  the solution to equation
\be\label{eig}
\vartheta l\eta\  \Phi_{k,1} =\ m \ \Phi_{k,1}\ , 
\ee
where $m$ is an  eigenvalue.
We look for the following most general solution
\be
\Phi_{k,1} = (\vartheta_i +c (\vartheta  l)_{ i} )\Phi_{k,1}^i\ , 
\ee
where $\Phi^i$ does not depend on $\vartheta$
and $c$ should be determined.
Making use of 
\be
\{(\vartheta l)_j,(l\eta)^i\}
=\frac{1}{2}(l^2+2\vartheta l\eta)\delta_j^i
+(\eta\vartheta-1)l^i{}_j
\ee
and $\eta^j \Phi_{k,1}^i=0$ we get
\be
(\vartheta l\eta)\Phi_{k,1}
= \Bigl((1+c)(\vartheta l)_i+ \frac{k(k+2)}{4}c\vartheta_i
\Bigr)\Phi_{k,1}^i\,.
\ee
{}From (\ref{eig}) we find then the equations
\be
\frac{k(k+2)}{4}c = m\,,
\qquad
(1+c) = m c\,,
\ee
which are solved by 
\be\label{mu1}
m^{(1)} = -\frac{k}{2}\,,
\qquad
c^{(1)} = -\frac{2}{k+2}\,;
\qquad
m^{(2)} = \frac{k+2}{2}\,,
\qquad
c^{(2)} = -\frac{2}{k}\,.
\ee
Thus we have the  two solution and the  two eigenvectors
\be
\Phi_{k,1}^{(1)} = 
(\vartheta_i -\frac{2}{k+2} (\vartheta l)_i)\Phi_{k,1}^i,
\qquad
\Phi_{k,1}^{(2)} = (\vartheta_i -\frac{2}{k} (\vartheta l)_i)
\Phi_{k,1}^i\,. 
\ee
Taking into account the relation
\be
2l^2 \Phi_{k,1}^{(1,2)} = k(k+2)\Phi_{k,1}^{(1,2)}\,,
\qquad
(\eta\vartheta-1) \Phi_{k,1}^{(1,2)} = 0\,,
\ee
and the eigenvalues  $m^{(1)}$ and $m^{(2)}$ 
of   $\vartheta l\eta$  given in (\ref{mu1})
we get the following eigenvalues of the operator $X$
\be
X \Phi_{k,1}^{(1)} = k^2 \Phi_{k,1}^{(1)}\,,
\qquad
X \Phi_{k,1}^{(2)} = (k+2)^2 \Phi_{k,1}^{(2)}\,.
\ee


\appendix{Superstring action}


The standard kinetic term of superstring action 
in $AdS_3\times S^3$  \cite{ad,st,ri}
\begin{equation}\label{lkin0}
{\cal L}_{kin}
=-\frac{1}{2}\sqrt{g}g^{\vm\vn}
\Bigl(\hat{L}_\vm^A\hat{L}_\vn^A
+L_\vm^\apr L_\vn^\apr\Bigr) \ ,
\end{equation}
can be rewritten in conformal algebra 
notation  as \cite{MT3}
\begin{equation}\label{lkin1}
{\cal L}_{kin}
=-\frac{1}{2}\sqrt{g}g^{\vm\vn}
\Bigl(\hat{L}_\vm^a\hat{L}_\vn^a
+L_{\ssmD\vm}L_{\ssmD\vn}
+L_\vm^\apr L_\vn^\apr\Bigr) \ ,
\end{equation}
where the Cartan 1-forms 
\begin{equation}\label{hll}
\hat{L}^a\equiv L_\ssmP^a-\frac{1}{2}L_\ssmK^a\,,
\qquad
L^\apr\equiv -\frac{\rm i}{2}(\sigma^\apr)^i{}_j L^j{}_i
+\frac{\rm i}{2}(\sigma^\apr)^i{}_j \tilde{L}^j{}_i\,,
\end{equation}
in the light-cone basis are defined by
\begin{eqnarray}\label{carfor}
&&
G^{-1}dG
=
L_\ssmP^aP^a+L_\ssmK^aK^a+L_\ssmD D  
+L^{-+}J^{+-} + L^i{}_j J^j{}_i+ \tilde{L}^i{}_j \tilde{J}^j{}_i
\\ &&
+\  L_\ssmQ^{-i}Q_i^+ + L_{\ssmQ\, i}^-Q^{+i}
+ L_\ssmQ^{+i}Q_i^- + L_{\ssmQ\,i}^+ Q^{-i}
+\ L_\ssmS^{-i}S_i^+ + L_{\ssmS\, i}^- S^{+i}
+ L_\ssmS^{+i} S_i^- + L_{\ssmS\,i}^+ S^{-i}\,,
\nonumber
\end{eqnarray}
where the 
generators are taken in the basis described in Appendix B.
To represent the Cartan 1-forms in terms of the
even and odd coordinate      fields  we shall
start with the following supercoset representative    
\begin{eqnarray}
\label{kilgau}
G&=& \exp(x^aP^a +\theta^{-i} Q_i^+ 
+\theta_i^-Q^{+i}+\theta^{+i} Q_i^-
+\theta_i^+Q^{-i})
\\
& &\times 
\exp(\eta^{-i}S_i^+ + \eta_i^- S^{+i}+\eta^{+i}S_i^- 
+\eta_i^+S^{-i})
\, g_y\ g_\phi \ ,
\nonumber
\end{eqnarray}
where
$g_\phi$ and $g_y$ depend on the  radial $AdS_3$ coordinate $\phi$
and $S^3$ coordinates $y^\apr$ respectively:
\be\label{gy}
g_\phi \equiv \exp(\phi D)\ , 
\qquad
g_y\equiv \exp(y^i{}_j(J^j{}_i-\tilde{J}^j{}_i))\,,
\qquad
 y^i{}_j\equiv \frac{{\rm i}}{2}(\sigma^A)^i{}_j y^\apr \ .
\ee
Choosing the parametrization of the
 coset representative in the  form (\ref{kilgau})
 corresponds to what is   usually    referred
to as   ``Killing gauge"   in superspace.
Eq. \rf{carfor} 
 provides the  definition of the Cartan forms
 in the light-cone basis.
 Let us   further specify  them by setting
 to zero some of the fermionic coordinates
  which  corresponds to fixing
   a  particular  $\kappa$-symmetry gauge.
Namely, we shall
 fix the $\kappa$-symmetry by putting to
 zero all the  Grassmann  coordinates which carry
positive $J^{+-}$ charge:
$\theta^{+i}=\theta_i^+=\eta^{+i}=\eta_i^+=0$.
To simplify the notation   in what follows we shall set 
$\theta^i\equiv \theta^{-i}$,
$\theta_i\equiv \theta_i^-$,
$\eta^i\equiv \eta^{-i}$,
$\eta_i\equiv \eta_i^-$.
Note that since 
$S^{+i}$ and $Q^{+i}$ transform in the fundamental representations
of $su(2)$ and $\widetilde{su}(2)$ the
corresponding  fermionic coordinates $\eta$'s  and $\theta$'s
also transform
in the fundamental representation of $su(2)$ and $\widetilde{su}(2)$.
As a result,
the  $\kappa$-symmetry fixed  form of the
coset representative (\ref{kilgau})  is
\be\label{kapfixG}
G_{g.f.} 
=\exp(x^aP^a +\theta^i Q_i^+ +\theta_iQ^{+i}) \ 
\exp(\eta^iS_i^+ + \eta_iS^{+i})
\ g_y\ g_\phi \ .
\ee
Plugging  $G_{g.f.}$  into  (\ref{carfor}) we get
the $\kappa$-symmetry gauge fixed expressions for the Cartan
1-forms
\begin{eqnarray}
\label{carfor1}
&&
L_\ssmP^+ = e^\phi dx^+   \ , \ \ \ \   \
L_\ssmP^-=e^\phi (dx^-
- \frac{{\rm i}}{2}\tilde{\theta}^i\tilde{d\theta}_i
-\frac{{\rm i}}{2}\tilde{\theta}_i \tilde{d{\theta^i}})   \ ,
\\
&&
L_\ssmK^-=e^{-\phi}\bigl[\frac{1}{4}(\tilde{\eta}^2)^2 dx^+
+\frac{{\rm i}}{2}\tilde{\eta}^i \tilde{d\eta}_i
+\frac{{\rm i}}{2}\tilde{\eta}_i \tilde{d{\eta^i}}\bigr]   \ ,
\qquad
L_\ssmD=d\phi \ ,  
\\
&&
L^i{}_j= (dUU^{-1})^i{}_j+{\rm i}(\tilde{\eta}^i\tilde{\eta}_j
-\frac{1}{2}\tilde{\eta}^2\delta_j^i)dx^+   \ ,  
\ \  \ \ \ \ \widetilde{L}^i{}_j= (d\tilde{U}\tilde{U}^{-1})^i{}_j
 \ ,  \label{carforij}
\\
&&
L_\ssmQ^{-i}=e^{\phi/2}\tilde{d\theta}^i,
\quad,
L_{\ssmQ i}^-=e^{\phi/2}\tilde{d\theta}_i\,,
\quad
L_\ssmQ^{+i}=-{\rm i}e^{\phi/2}\tilde{\eta}^i dx^+,
\quad
L_{\ssmQ i}^+={\rm i}e^{\phi/2}\tilde{\eta}_i dx^+,
\qquad 
\\
&&
L_\ssmS^{-i}=e^{-\phi/2}(\tilde{d\eta}^i
+\frac{{\rm i}}{2}\tilde{\eta}^2\tilde{\eta}^i dx^+)
\ , \ \ \ \ \ \ \
L_{\ssmS i}^- = e^{-\phi/2}(\tilde{d\eta}_i
-\frac{{\rm i}}{2}\tilde{\eta}^2\tilde{\eta}_i dx^+)
\ , \label{carfor2}
\end{eqnarray}
with all  the remaining forms  equal to zero.
We  have  introduced the notation
\begin{equation}\label{tilthe}\widetilde{\eta}^i \equiv  U^i{}_j \eta^j\,,
\qquad\widetilde{\eta}_i \equiv  \eta_j (U^{-1})^j{}_i  \ ,
\qquad\widetilde{\theta}^i \equiv  \tilde{U}^i{}_j \theta^j\,,
\qquad\widetilde{\theta}_i \equiv  \theta_j (\tilde{U}^{-1})^j{}_i  \ ,
\end{equation}
\begin{equation}\label{tildthe}\widetilde{d\eta}^i \equiv U^i{}_j d\eta^j\,,
\qquad\widetilde{d\eta}_i \equiv  d\eta_j (U^{-1})^j{}_i      \  ,
\qquad\widetilde{d{\theta^i}}\equiv \tilde{U}^i{}_j d\theta^j\,,
\qquad\widetilde{d\theta}_i \equiv  d\theta_j (\tilde{U}^{-1})^j{}_i      \  .
\end{equation}
The fact that $\theta^i$ and $\eta^i$ are rotated by $\widetilde{SU}(2)$ and 
$SU(2)$ is related to   the presence of the 
matrices $\tilde{U}\in \widetilde{SU}(2)$
and $U\in SU(2)$ in the definition of $\tilde{\theta}^i$ and $\tilde{\eta}^i$.
These matrices defined by (\ref{uuu}) 
can be written  explicitly as
\be\label{uuuu}
U=\cos{|y|\over 2 } +{\rm i}\sigma^\apr n^\apr \sin{|y|\over 2 }\,,
\qquad\widetilde{U}=\cos{|y|\over 2 } -{\rm i}\sigma^\apr n^\apr
 \sin{|y|\over 2 }\ , 
\ee
where $|y|$ and $n^\apr$ are given by (\ref{b2}). 

The $S^3$  components $L^\apr$  of the 
Cartan forms defined by
(\ref{hll}) can be written  in the following equivalent ways
\be\label{la12}
L^\apr = e^\apr 
-\frac{1}{2}\tilde{\eta}_i(\sigma^\apr)^i{}_j\tilde{\eta}^jdx^+\ 
\ , \ \ \  \qquad
L^\apr = e^\apr_\sca(dy^\sca 
+{\rm i}\eta_i (V^\sca)^i{}_j\eta^jdx^+)
\ee
where $e^\apr_\sca$ and $(V^\sca)^i{}_j$ are defined by
(\ref{b1}) and (\ref{b5})
and we used the  relation
\be
e^\sca_\apr (U^\dagger \sigma^\apr U)^i{}_j 
= -2{\rm i}(V^\sca)^i{}_j\,,
\qquad
e^\sca_\apr e_\sca^\bpr =\delta_\apr^\bpr\,.
\ee
Plugging the Cartan 1-forms into  the kinetic part of the 
 string Lagrangian 
(\ref{lkin1}) we get
\begin{eqnarray}
{\cal L}_{kin}
&=&
-\sqrt{g}g^{\vm\vn}
\Bigl[e^{2\phi}\partial_\vm x^+ \partial_\vn x ^- 
+\frac{1}{2}\partial_\vm\phi\partial_\vn \phi
+\frac{1}{2}L_\vm^\apr L_\vn^\apr\Bigr]
\\
&+&\sqrt{g}g^{\vm\vn}\partial_\vm 
x^+\Bigl[(e^{2\phi}(\frac{{\rm i}}{2}\theta^i\partial_\vn\theta_i
+\frac{{\rm i}}{2}\theta_i\partial_\vn\theta^i)
+\frac{{\rm i}}{4}\eta^i\partial_\vn\eta_i
+\frac{{\rm i}}{4}\eta_i\partial_\vn\eta^i
+\frac{1}{8}(\eta^2)^2\partial_\vn x^+\Bigr]\,.
\nonumber
\end{eqnarray}
In order to get the 
action in the  Killing parametrization 
one needs to use $L^\apr$ given by the second expression in
(\ref{la12}) and then make the rescalings
\be
\eta^i \rightarrow \sqrt{2}e^\phi \eta^i,\ 
\quad
\eta_i \rightarrow \sqrt{2}e^\phi\eta_i\,,\ \  
\quad
x^a\rightarrow -x ^a\,.
\ee 
The action in the  WZ parametrization 
is found by using 
$L^\apr$ given by the first expression in (\ref{la12})
and after the transformation 
\be
\theta^i \rightarrow  (\tilde{U}^{-1})^i{}_j\theta^i,
\quad
\theta_i \rightarrow \theta_j \tilde{U}^j{}_i\,,
\quad
\eta^i \rightarrow \sqrt{2}e^\phi (U^{-1})^i{}_j\eta^j,
\quad
\eta_i \rightarrow \sqrt{2}e^\phi\eta_j U^j{}_i\,,
\quad
x^a\rightarrow -x ^a\ , 
\ee 
and use of the  Fierz rearrangement rule
$(\eta_i (\sigma^\apr)^i{}_j \eta^j)^2 = -3(\eta^2)^2$.

The P-odd WZ part of the  covariant string Lagrangian
 ${\cal L}_{WZ}$  (see e.g. \ci{ad,st,ri})
 takes the following form 
  in the light-cone gauge
\be
{\cal L}_{WZ}
=-\frac{\rm i}{\sqrt{2}}
\epsilon^{\vm\vn}L_{\ssmQ \vm}^{+i}\Csp_{ij} L_{\ssmQ \vn}^{-j}
+h.c.\ . 
\ee
Plugging in  the  expressions for 
the  Cartan 1-forms and making the rescalings
given above we get ${\cal L}_{WZ}$ in  the Killing
parametrization (\ref{actwz3}) (see also (\ref{actwz6})) and in 
the WZ
parametrization (see the $\epsilon^{\mu\nu}$ term in (\ref{ff})).


\end{document}